\documentclass[onecolumn,notitlepage,a4paper,superscriptaddress,nofootinbib,longbibliography]{revtex4-2}

\usepackage[english]{babel}
\usepackage[T1]{fontenc}
\usepackage[utf8]{inputenc}
\usepackage{amsmath,accents}
\usepackage{amsfonts,color}
\usepackage{amssymb,float}
\usepackage{mathptmx}
\usepackage{mathrsfs}

\usepackage{graphicx}
\usepackage{subfigure}
\usepackage{color}
\usepackage{subcaption}

\usepackage[pdftoolbar = true, pdfstartview = FitH, pdfmenubar = true]{hyperref}
\hypersetup{
    colorlinks=true,
    linkcolor=red,
    filecolor=magenta,      
   citecolor=cyan
}

\newcommand{\lc}[1]{\accentset{\circ}{#1}}

\newcommand{\be}{\begin{equation}}
\newcommand{\ee}{\end{equation}}
\newcommand{\bea}{\begin{eqnarray}}
\newcommand{\eea}{\end{eqnarray}}

\begin{document}

\title{Exploring the stability of \texorpdfstring{$f(Q)$}{} cosmology near general relativity limit with different connections}

\author{ Mar\'ia-Jos\'e Guzm\'an}
\email{mjguzman@ut.ee}
\affiliation{Institute of Physics, University of Tartu, W.\ Ostwaldi 1, 50411 Tartu, Estonia}
\author{Laur Järv} 
\email{laur.jarv@ut.ee}
\affiliation{Institute of Physics, University of Tartu, W.\ Ostwaldi 1, 50411 Tartu, Estonia}
\author{Laxmipriya Pati}
\email{laxmipriya.pati@ut.ee}
\thanks{corresponding author}
\affiliation{Institute of Physics, University of Tartu, W.\ Ostwaldi 1, 50411 Tartu, Estonia}

\begin{abstract}
In this work, we study cosmological spacetime configurations in $f(Q)$ gravity with nonvanishing symmetric teleparallel connections. It is known that the spatially flat, homogeneous and isotropic connections can be classified into three sets. Focusing on two of those, we explore the stability of cosmological background evolution near the general relativity regime across radiation, matter, dark energy, and geometric dark energy dominated eras. Our results show that for the standard connection set 1 the general relativity regime can be realized in two ways and both exhibit stable behavior throughout all evolutionary epochs. Conversely, for the alternative connection set 2 the trivial general relativity limit is stable, while the nontrivial option exhibits stability during the radiation era and marginal stability during the matter era, but for the dark energy and geometric dark energy eras our results are inconclusive. Furthermore, we discuss the general conditions on the function $f(Q)$ that physically viable models should obey, and point out that for a generic $f(Q)$ the alternative connection sets 2 and 3 are prone to trigger a sudden singularity. This can happen even near the otherwise good looking general relativity regime, as we demonstrate by an explicit numerical example. Hence the alternative configurations could be problematic already on the background level.
\end{abstract}
\maketitle

\section{Introduction}
General relativity (GR) is our most successful theory to describe the Universe at cosmological scales. It is the foundation of the $\Lambda$CDM cosmological model, which  incorporates a cosmological constant, baryonic and cold dark matter. This model describes various phenomena observed in the universe such as the existence and structure of the cosmic microwave background, the large-scale structure in the distribution of galaxies, and the accelerating expansion of the universe observed by distant galaxies and supernovae. Nonetheless, the $\Lambda$CDM faces problems such as the Hubble tension produced by the difference in the measurements of the Hubble parameter by using different methods, or the nature of the dark energy traditionally associated to the cosmological constant, but challenged by recent hints of a varying equation of state \cite{Perivolaropoulos:2021jda,DESI:2024mwx}.

One of the ways of proposing alternatives for going beyond general relativity relies on relaxing the assumptions leading to the choice of the Levi-Civita connection. These are the imposition of a vanishing torsion and nonmetricity on the most general linear connection. Remarkably, the understanding of gravity as the curvature of spacetime has two additional alternative geometrical descriptions given by the torsion and nonmetricity tensors. In the framework of the teleparallel equivalent of general relativity (TEGR) \cite{Aldrovandi:2013wha}, the affine connection has torsion but vanishing curvature and nonmetricity. Instead, in the symmetric teleparallel equivalent of general relativity (STEGR) \cite{Nester:1998mp,Adak:2006rx} the curvature and the torsion tensor vanish, and the main dynamical object that carries the gravitational interaction is the nonmetricity tensor. It is then possible to obtain the dynamics of Einstein's equations not only from the Levi-Civita curvature scalar $\mathring{R}$ but also from the torsion scalar $T$ or nonmetricity scalar $Q$ built up from the torsion and nonmetricity tensors, respectively. These scalars are related to $\mathring{R}$ by boundary terms $B_T$ and $B_Q$ in such a way that $\mathring{R} = -T + B_T$ and $\mathring{R} = -Q + B_Q$. Both boundary terms depend on the teleparallel connection, but for TEGR and STEGR these terms drop out of the equations of motion \cite{BeltranJimenez:2019tjy,Bahamonde:2021gfp,Capozziello:2022zzh,Jarv:2023anw}. This is better seen by replacing the curvature scalar by its torsion or nonmetricity scalar counterparts in the Einstein-Hilbert action, then the boundary term is integrated out, it does not contribute to the equations of motion, and we obtain the same dynamics as in GR, given that the matter Lagrangian is left unchanged and matter only couples to the metric (or Levi-Civita connection). Therefore, the connection remains a pure gauge choice, unless considering disciplines where it could have repercussions such as numerical relativity \cite{Capozziello:2021pcg,Pati:2022nwi,Guzman:2023oyl} or physical problems where the boundary terms become relevant such as black hole energy or entropy \cite{BeltranJimenez:2021kpj,Gomes:2022vrc,Koivisto:2022oyt}. The metric teleparallel and symmetric teleparallel paradigms represented by TEGR and STEGR can be contained into a generalized parent theory known as general teleparallel equivalent of general relativity (GTEGR), that considers both torsion and nonmetricity and its Lagrangian is built from the torsion and nonmetricity scalars plus mixed terms combining the torsion and nonmetricity tensors \cite{BeltranJimenez:2019odq,Bajardi:2024qbi}.

In the same way that the Einstein-Hilbert Lagrangian can be extended by considering a nonlinear function $f(R)$, e.g.\ to include quantum corrections to the GR action or improve its renormalizability \cite{CANTATA:2021asi}, it is possible to consider as a starting point the torsion $T$  and nonmetricity $Q$ scalars, giving rise to $f(T)$ \cite{Ferraro:2006jd,Bengochea:2008gz} and $f(Q)$ \cite{BeltranJimenez:2017tkd,BeltranJimenez:2018vdo,BeltranJimenez:2019tme} gravities, motivated by similar considerations as in GR \cite{CANTATA:2021asi}. Noticeably, these extensions do not deliver the same dynamics as in $f(R)$ gravity, since now the boundary terms are encapsulated inside the function $f$, giving rise to nontrivial dynamics that  manifests in the field equations. These  dynamics can be understood from the point of view of the connection that partially appears in the boundary terms $B_T$ and $B_Q$. Since the boundary terms do not drop out of the equations of motion for $f(T)$ and $f(Q)$, they grant dynamics to some parts of the connection, which now have their own equation. This fact  makes nontrivial the task of finding solutions for the combined system of equations for the metric and the connection. One procedure that has been fruitful in easing this task, is to impose an ansatz for the connection which obeys the same set of spacetime symmetries as the metric \cite{Hohmann:2019nat,Jarv:2019ctf}. In the case of a Friedmann-Lemaître-Robertson-Walker (FLRW) cosmological spacetime, the metric teleparallel connection, which only has torsion, provides one family of connections for a cosmology with a flat spatial slice, and two families for a curved spatial slice \cite{Hohmann:2020zre,Coley:2022qug}. Contrary to it, the symmetric teleparallel connection with only nonvanishing nonmetricity, has three different options that are compatible with spatial flatness and one for spatial curvature, where all cases show up an extra free function of time in the connection \cite{Hohmann:2021ast,DAmbrosio:2021pnd}. The diversity of families involves a distinct dynamics for each one, that we would like to study in this work. 

We will consider the cosmological evolution equations of $f(Q)$ gravity in the covariant formalism, where the theory exhibits general diffeomorphism invariance under simultaneous diffeomorphisms on both the metric and the connection. Our starting point is the FLRW metric in spherical coordinates, that is paired up with a nontrivial connection leading to four distinct solution branches \cite{Hohmann:2021ast, DAmbrosio:2021pnd}.
Following Refs.\ \cite{Zhao:2021zab,Hohmann:2021ast, DAmbrosio:2021pnd}, the solutions are organized into two groups, where the first group consists of three branches with different connections and with zero spatial curvature, while the second group exhibits nonzero spatial curvature. One of the branches of the first group has background cosmological equations of motion that coincide with the equations for $f(T)$ gravity, which have been extensively studied in the literature. However, much less research has been conducted within the nontrivial branches of the connection. There have been some studies considering particular solutions and their properties \cite{Dimakis:2022rkd,Dimakis:2022wkj,Paliathanasis:2023ngs}, phase space features \cite{Shabani:2023nvm,Paliathanasis:2023nkb,Shabani:2023xfn}, and constraints from observational data \cite{Subramaniam:2023old,Shi:2023kvu,Yang:2024tkw}, all for certain specific classes of models of $f(Q)$. On the one hand there are claims that these alternative connections can perform better in fitting the observational Hubble data than the $\Lambda$CDM and other models \cite{Yang:2024tkw}. On the other hand, it was shown in the context of scalar-tensor symmetric teleparallel gravity that the alternative FLRW connections are prone to trigger a sudden singularity for a wide range of initial conditions \cite{Jarv:2023sbp}, while examples of divergent cosmological behavior have been subsequently also noted by other authors \cite{Paliathanasis:2023gfq,Yang:2024tkw}.

These issues link to a deeper puzzle which is that $f(Q)$ theories seem to exhibit a background dependent number of dynamical degrees of freedom at the linear perturbation level \cite{BeltranJimenez:2019tme,BeltranJimenez:2021auj,Gomes:2023tur,Heisenberg:2023wgk,Zhao:2024kri}. This can point to the strong coupling problem, which can be understood as a mismatch in the number of degrees of freedom appearing in the perturbations around most common backgrounds, \textit{vis-à-vis} the number of degrees of freedom expected from the Hamiltonian analysis. Oddly, in the case of $f(Q)$ gravity the researchers have not reached a consensus of how many degrees of freedom does the Hamiltonian analysis predict \cite{Hu:2022anq,DAmbrosio:2023asf,Tomonari:2023wcs}. If following the case of $f(T)$ gravity where discrepancies on the counting also appear, this could be attributed to the nonlinearities on the constraints of the theory that produce a matrix of Poisson brackets among constraints with variable rank, rendering an utterly complicated analysis \cite{Golovnev:2020nln,Golovnev:2020zpv}. From another perspective, a recent investigation of the metric cosmological perturbations indicated the presence of a ghost degree of freedom in the ultraviolet regime within the spacetime configuration given by one of the alternative connections \cite{Gomes:2023tur}. This would have serious implications to the viability of $f(Q)$ gravity in general, and raises the importance of a better understanding of how stable are the cosmological configurations with the alternative connections already in the background level. If the alternative FLRW backgrounds could be deemed a mathematical artifact otherwise unsuitable for serious physical applications, then the presence of ghost modes implied by them would perhaps not be so detrimental. 
 
Since even in the cosmology of general relativity there are singularities which are not considered to be totally alarming  (like the Big Bang or Big Crunch) we need to set our focus more clearly. Despite some shortcomings, the $\Lambda$CDM model in general relativity provides a description of the cosmic history in robust agreement with the data, and most likely the improvements offered by modified gravity will be in a form of modest deviations from $\Lambda$CDM. Therefore, in the present work we will investigate the stability of $f(Q)$ solutions which are close to the GR regime in the radiation, dust matter, and cosmological constant dominated eras, using the methods developed earlier in the context of scalar-tensor theories  \cite{Damour:1993id,Serna:1995pi,Mimoso:1998dn,Santiago:1998ae,Serna:2002fj,Jarv:2010zc,Jarv:2010xm,Jarv:2011sm,Jarv:2015kga,Dutta:2020uha,Jarv:2015odu,Jarv:2023sbp}.
The expectation is that in a viable $f(Q)$ model all solutions that initially evolve quite similarly to their GR counterparts should remain to do so, which guarantees that the succession from radiation to dust matter to dark energy domination follows naturally from the different scaling laws of the respective energy densities. If the solutions in a $f(Q)$ model get diverted from this scenario, then the model is most likely not viable to match the observations. If the solutions hit a sudden singularity in that process, then the model can be called unstable and neglected on more fundamental grounds. In this respect we analyze in detail the first and second FLRW connection branches, to see how the first (standard, or trivial) case compares to the second (alternative) case which involves an extra free function and where a ghost was reported in Ref.\ \cite{Gomes:2023tur}.

This work is organized as follows. Sec.\ \ref{Symmetric teleparallelism} provides a foundational overview of the symmetric teleparallel framework, the symmetric teleparallel equivalent of general relativity and its nonlinear $f(Q)$ extension, together with their equations of motion. Sec.\ \ref{Cosmology of spatially homogeneous and isotropic field configurations} explores cosmological homogeneous and isotropic backgrounds of the three possible connection sets in $f(Q)$ gravity, and the number of independent dynamical functions defined by the equations of motion. 
We establish the conditions under which the equations of motion reduce to GR and present the solution for the GR background in Sec.\ \ref{General Relativity Limit}. In Sec.\ \ref{Connection set I} we define the background evolution for connection set 1 and study its stability under small perturbation of the dynamical variables. The same analysis is performed in Sec. \ref{Connection set II} for connection set 2. In both these two sections the study focused on the stability of four cosmological regimes: matter, radiation, dark energy, and geometric dark energy dominated eras.
In Sec.\ \ref{Numerical example} we present a numerical example in order to better understand the marginal stability found for connection set 2, and encounter a sudden singularity. Finally, Sec.\ \ref{Conclusions} presents a concise summary and discussion of the results.

\section{Symmetric teleparallel geometry and gravity}
\label{Symmetric teleparallelism}

\subsection{Symmetric teleparallel geometry}

We begin by introducing a non-Riemannian affine manifold $\mathcal{M}$ denoted as $\left(\mathcal{M},g_{\mu\nu},\tilde{\Gamma}^{\alpha}{}_{\mu\nu}\right)$, where $g_{\mu\nu}$ is a metric tensor with signature $\left(-,+,+,+\right)$ and $\tilde{\Gamma}^{\alpha}{}_{\mu\nu}$ represents the general affine connection with 64 independent components, that can be decomposed into three pieces,
\begin{equation}
\label{Connection decomposition}
\tilde{\Gamma}{}^{\lambda}{}_{\mu\nu} = \lc{\Gamma}{}^{\lambda}{}_{\mu\nu} + K^{\lambda}{}_{\mu\nu}  + L^{\lambda}{}_{\mu\nu} \,,
\end{equation}
whereas the Levi-Civita connection $\lc{\Gamma}{}^{\lambda}{}_{\mu\nu}$ of the metric is
\begin{equation}
\label{LeviCivita}
 \lc{\Gamma}^{\lambda}{}_{\mu \nu} \equiv \frac{1}{2} g^{\lambda \beta} \left( \partial_{\mu} g_{\beta\nu} + \partial_{\nu} g_{\beta\mu} - \partial_{\beta} g_{\mu\nu} \right)= \lc{\Gamma}^{\lambda}{}_{\nu \mu}\,,
\end{equation}
the contortion tensor is
\begin{align}
    K^{\lambda}{}_{\mu\nu} \equiv \frac{1}{2} g^{\lambda \beta} \left( -T_{\mu\beta\nu}-T_{\nu\beta\mu} +T_{\beta\mu\nu} \right) =-\,K_{\mu}{}^{\lambda}{}_{\nu}\,,
\end{align}
and the disformation tensor defined as
\begin{equation}
\label{Disformation}
L^{\lambda}{}_{\mu\nu} \equiv \frac{1}{2} g^{\lambda \beta} \left( -Q_{\mu \beta\nu}-Q_{\nu \beta\mu}+Q_{\beta \mu \nu} \right) = L^{\lambda}{}_{\nu\mu}  \,.
\end{equation}
By construction, both the Levi-Civita connection and the disformation tensor are symmetric.
The contortion tensor is constructed from the torsion tensor which captures the antisymmetric part of the connection as
\begin{align}
    \label{TorsionTensor}
        T^{\lambda}{}_{\mu\nu} \equiv \tilde{\Gamma}{}^{\lambda}{}_{\mu\nu}-\tilde{\Gamma}{}^{\lambda}{}_{\nu\mu}\,,
\end{align}
while the disformation tensor built from the nonmetricity tensor is symmetric in its last two indices
\begin{equation}
\label{NonMetricityTensor}
Q_{\rho \mu \nu} \equiv \nabla_{\rho} g_{\mu\nu} = \partial_\rho g_{\mu\nu} - \tilde{\Gamma}{}^\beta{}_{\mu \rho} g_{\beta \nu} -  \tilde{\Gamma}{}^\beta{}_{\nu \rho } g_{\mu \beta}  \,,
\end{equation}
and it measures how the connection deviates from being metric-compatible.
The curvature tensor
\begin{align}
\label{CurvatureTensor}
    \tilde{R}^{\sigma}\,_{\rho\mu\nu} (\tilde{\Gamma}) \equiv \partial_{\mu}\tilde{\Gamma}{}^{\sigma}\,_{\nu\rho}-\partial_{\nu}\tilde{\Gamma}{}^{\sigma}\,_{\mu\rho}+\tilde{\Gamma}{}^{\sigma}\,_{\mu\lambda}\tilde{\Gamma}{}^{\lambda}\,_{\nu\rho}-\tilde{\Gamma}{}^{\sigma}\,_{\mu\lambda}\tilde{\Gamma}{}^{\lambda}\,_{\nu\rho}\,, 
\end{align}
torsion tensor and nonmetricity tensor are the key properties of a connection that characterizes parallel transport. Teleparallel geometry is defined by vanishing curvature \eqref{CurvatureTensor} since then the vectors maintain their direction in parallel transport. Vanishing torsion \eqref{TorsionTensor} implies that the affine connection is symmetric in the lower indices, while both vanishing  curvature \eqref{CurvatureTensor} and torsion \eqref{TorsionTensor} are known as symmetric teleparallel geometry.

Considering all the possible quadratic combinations of the nonmetricity tensor \eqref{NonMetricityTensor} allows the construction of the scalar quantity
\begin{equation}\label{Generalnonmetricity}
\mathbb{Q} = \alpha_{1}\,Q_{\lambda\mu\nu}Q^{\lambda\mu\nu}+\alpha_{2}Q_{\lambda\mu\nu}Q^{\mu\nu\lambda}+\alpha_{3}Q_{\mu}Q^{\mu} +\alpha_{4}\hat{Q}_{\mu}\hat{Q}^{\mu} + \alpha_{5}\,Q_{\mu}\hat{Q}^{\mu}\,, 
\end{equation}
where $\alpha_{i}$ are arbitrary real numbers and  $Q_{\mu}\equiv Q_{\mu\nu}{}^{\nu}\,$
, \text{and}
$\hat{Q}_{\mu} \equiv Q_{\nu\mu}{}^{\nu}\,$ are two independent traces arising from the symmetry properties of the nonmetricity tensor. With a specific set of the parameters
\begin{align}
\label{STEGR_parameter}
    \alpha_{1}= -\frac{1}{4}, \quad \quad \alpha_{2}=\frac{1}{2}, \quad \quad \alpha_{3}=\frac{1}{4}, \quad \quad \alpha_{4}=0, \quad \quad \alpha_{5}=-\frac{1}{2},
\end{align}
in equation \eqref{Generalnonmetricity}, the nonmetricity scalar is expressed as
\begin{equation}
\label{STEGRnonmetricity}
Q\equiv -\,\frac{1}{4}\,Q_{\lambda\mu\nu}Q^{\lambda\mu\nu}+\frac{1}{2}\,Q_{\lambda\mu\nu}Q^{\mu\nu\lambda}+\frac{1}{4}\,Q_{\mu}Q^{\mu}-\frac{1}{2}\,Q_{\mu}\hat{Q}^{\mu}\,.
\end{equation}
In the case of symmetric teleparallelism there is an interesting relation between the Levi-Civita curvature and the nonmetricity scalar
\begin{align} \label{geometricidentity}
    \lc{R}=  Q + \lc{\nabla}_{\mu}(\hat{Q}^{\mu}-Q^{\mu}) \,,
\end{align}
where $\lc{\nabla}_{\mu}$ is the covariant derivative with respect to the Levi-Civita connection. The second term of \eqref{geometricidentity} can be represented as 
\begin{align}
    \label{boundary_term}
    B_{Q}=\lc{\nabla}_{\mu}(\hat{Q}^{\mu}-Q^{\mu})
\end{align}
which is commonly referred to as the boundary term of STEGR. Considering the specific parameter set \eqref{STEGR_parameter} in equation \eqref{Generalnonmetricity} leads to the recovery of general relativity action up to the boundary term \eqref{geometricidentity} while due to the presence
of these arbitrary parameters in \eqref{Generalnonmetricity} the theory of the
general quadratic symmetric teleparallel Lagrangian \eqref{Generalnonmetricity}  itself is not equivalent to GR.

The boundary term \eqref{boundary_term} holds considerable significance in various areas of physics, including its connection to the entropy of black holes, Casimir effect or Hamiltonian formulation. Identity \eqref{geometricidentity} reveals that the Einstein Hilbert Lagrangian $\lc{R}$ 
yields the same field equations 
as the symmetric teleparallel Lagrangian $Q$, since the boundary term does not play any role when one acts the variational computation on the manifold. Nevertheless this boundary term actually should play a role when one considers a manifold with a boundary. So far these issues have not been considered much in the teleparallel context. 

\subsection{\texorpdfstring{$f(Q)$}{} gravity} \label{$f(Q)$ gravity}
Our starting point is the action for a nonlinear modification of the STEGR Lagrangian, that is the $f(Q)$ gravity action
\begin{equation}
\label{f(Q)Action}
S = \frac{1}{2\kappa^2} \int\mathrm{d}^4x \sqrt{-g} f(Q) + S_{\mathrm{m}}(g)\,.
\end{equation}
Here $Q$ is the nonmetricity scalar \eqref{STEGRnonmetricity} and $f$ is \textit{a priori} a free function which determines a particular theory within the general class. In the specific case $f(Q)=Q$ the action recovers the symmetric teleparallel equivalent of general relativity. The matter Lagrangian $S_{m}$ is considered to depend solely on the metric, hence hypermomentum is assumed to be zero. In our conventions $\kappa^2=8\pi G$ is a constant that relates to the Newtonian gravitational constant $G$ in the appropriate limit. Since the gravitational sector of the theory depends both on the metric and the symmetric teleparallel connection, varying the action with respect to these fundamental variables gives two equations. It is perhaps the most convenient to write the metric equation as  \cite{Lin:2021uqa}\footnote{ It may be worthwhile to note that our conventions differ from those in the review \cite{Hohmann:2021ast} regarding the sign of the nonmetricity scalar $Q$. To achieve consistency, one has to flip the sign of $f$ and $f_{QQ}$, but not of $f_Q$.}
\begin{subequations}
\label{eq: fieldequationf(Q)}
   \begin{align}
    \label{metricfieldeq}
        f_Q\lc{G}_{\mu\nu} +2f_{QQ} P^\lambda{}_{\mu\nu} \partial_\lambda Q + \frac{1}{2} g_{\mu\nu}\left(Q f_Q-f(Q)\right) = \kappa^2\mathcal{T}_{\mu\nu}\,,
    \end{align}
and the connection equation as \cite{BeltranJimenez:2017tkd}
    \begin{align}
      \label{connectionfieldeq}
        \nabla_\mu \nabla_\nu \left(\sqrt{-g}f_Q P^{\mu\nu}{}_\alpha\right) = 0\,.
   \end{align}
\end{subequations}
Here we have defined the matter energy momentum tensor as $\mathcal{T}_{\mu\nu}$, $\lc{G}_{\mu\nu}$ is the Einstein tensor computed from the Levi-Civita connection, while the superpotential tensor (also known as nonmetricity conjugate)   $P^\lambda{}_{\mu\nu}$ is given in terms of the nonmetricity tensor as
\begin{align}
    P^\lambda{}_{\mu\nu}=-\,\frac{1}{4}Q^{\lambda}{}_{\mu\nu}+\frac{1}{2}Q_{(\mu}{}^\lambda{}_{\nu)}+\frac{1}{4}g_{\mu\nu}Q^\lambda-\frac{1}{4}(g_{\mu\nu}\hat{Q}^\lambda+\delta^\lambda{}_{(\mu}Q_{\nu)}) \,.
\end{align}
The derivatives of $f$ are denoted as $f_Q=\frac{d f}{dQ}$, etc.

It has already been pointed out that the equations \eqref{eq: fieldequationf(Q)} are related to each other \cite{Jarv:2018bgs}. 
As the matter Lagrangian is not coupled to the connection, the energy momentum tensor is covariantly conserved, $\lc{\nabla}_{\mu}T^{\mu\nu}=0$. Then, performing the Levi-Civita covariant derivative of \eqref{metricfieldeq} as well as using a Bianchi identity, one can exactly arrive the connection equation \eqref{connectionfieldeq} \cite{Heisenberg:2023lru}. In other words, the connection equation is not independent, but rather a consequence of the metric field equation after using the Bianchi identity.

In the case $f(Q)=Q$ the metric equations \eqref{metricfieldeq} reduce to  Einstein's equations as in general relativity, while the connection equation \eqref{connectionfieldeq} becomes a geometric identity that is trivially satisfied. More interestingly, if $Q$ is constant, then $f$ and $f_Q$ are also constant, and the metric equations reduce to Einstein's equations with a cosmological constant term and rescaled gravitational constant, while the connection equation is again identically satisfied. Therefore, any metric which is a solution of GR or GR$+\Lambda$ is trivially also a solution of $f(Q)$ gravity for any $f$, and we just need to tune the symmetric teleparallel connection so that the nonmetricity scalar $Q$ is zero or constant with appropriate value. On the other hand, since the general equations \eqref{eq: fieldequationf(Q)} contain extra terms, the repertoire of solutions in $f(Q)$ gravity can be much richer than in GR or GR$+\Lambda$.

In contrast to the $f(R)$ theory of gravity with its fourth order equations of motion, the nonlinear extension of symmetric teleparallel gravity into $f(Q)$ theory yields only second order equations, similar to $f(T)$. However, the number of propagating degrees of freedom in $f(Q)$ remains an open question \cite{Tomonari:2023wcs,Hu:2022anq, Hu:2023gui,Heisenberg:2023lru,Gomes:2023tur} and likely differs from $f(R)$. This highlights the distinct dynamical behavior arising from nonlinear modifications of curvature, torsion, and nonmetricity formulation based gravity.

\subsection{General stability considerations of \texorpdfstring{$f(Q)$}{} gravity}
\label{sec: general conditions of stability}

It has not been extensively analyzed in the literature what are the conditions that the function $f$ in the field equations \eqref{eq: fieldequationf(Q)} should meet in order to produce physically viable models, which motivates us to shortly discuss it.
First, if $f_Q$ is negative, then the effective gravitational constant $\frac{\kappa^2}{f_Q}$ that sets how geometry responds to matter energy-momentum in Eq.\  \eqref{metricfieldeq} is also negative and the situation may be regarded as an anti-gravity. The borderline case of vanishing $f_Q$ implies infinite gravitational ``constant'' and corresponds to a singularity. Hence physically meaningful models should obey the requirement
\begin{align}
\label{eq: condition f_Q>0}
    f_Q > 0 \,.
\end{align}

Next, the third term on the right hand side (rhs) of \eqref{metricfieldeq} can be read as a variable cosmological constant. Generally, a negative cosmological constant drives the universe to collapse, which might be preferable to avoid. This requirement is not absolutely imperative, as reasonable solutions with negative cosmological constant can also exist. Analogously, an infinite cosmological constant would not make a long lasting happy universe, as in general everything would explode. These expectations can be written down as
\begin{align}
\label{eq: condition Q f_Q -f}
    0 \leq Q f_Q - f < \infty \,.
\end{align}
It probably also makes sense to ask that the derivative of the third term on the rhs of \eqref{metricfieldeq} would not become infinite. In the scalar-tensor picture that corresponds to the derivative of the scalar potential \cite{Jarv:2018bgs}, whose divergence means a singularity. We may write this expectation as an inverse,
\begin{align}
\label{eq: condtition 1/Qf_QQ}
    \frac{1}{Q f_{QQ}} \neq 0 \,.
\end{align}

The conditions outlined above are essentially functions of the nonmetricity scalar $Q$. The intricate nuance here is that the nonmetricity scalar may in principle change a sign during the evolution or inside the configuration of a given connection. Then it depends on the form of the function $f$ whether the conditions are immune to the change in the sign of $Q$ or not. For instance if $Q$ changes a sign, then the models of type $f(Q)=Q+Q^n$ can violate the condition \eqref{eq: condition f_Q>0} for even $n$ but not odd $n$. Therefore it seems we can not claim that a certain model $f(Q)$ is viable by itself, or that a certain connection is viable by itself, but these aspects must be investigated together which makes a comprehensive analysis complicated,  and perhaps could be carried out in terms of bifurcations and catastrophe theory \cite{Arnold}. It could be that some specific connection works without problems in specific models of $f(Q)$, but not in other models.

In the remainder of the paper, without a loss of generality, we will adopt the parametrization
\begin{equation}
\label{eq:param}
    f=Q+F(Q)
\end{equation}  
which explicitly brings out the nonlinear part of the model functions that deviates from general relativity.

\section{Cosmology of spatially homogeneous and isotropic field configurations \label{Cosmology of spatially homogeneous and isotropic field configurations}}
In order to derive the cosmological dynamics of symmetric teleparallel equations of motion \eqref{eq: fieldequationf(Q)}, one possible way is to assume the metric and symmetric teleparallel connection inherits the same cosmological symmetry of spacetime. 

\subsection{Cosmological symmetry}

We will obtain a suitable form for the connection and the metric by considering that a spatially homogeneous and isotropic cosmological spacetime is characterized by the Killing vectors of translations $\xi_{T_i}$ and rotations $\xi_{R_i}$ that can be written in spherical coordinates as
\begin{subequations}
\label{Killing_Vector}
\begin{align}
    \xi^{\mu}_{T_{x}} & = \left(0, \chi \sin(\theta) \cos(\phi), \frac{\chi}{r} \cos(\theta)\cos(\phi), -\frac{\chi}{r} \frac{\sin(\phi)}{\sin(\theta)} \right) \,, \\
    \xi^{\mu}_{T_{y}} & = \left(0, \chi \sin(\theta) \sin(\phi), \frac{\chi}{r} \cos(\theta)\sin(\phi), -\frac{\chi}{r} \frac{\cos(\phi)}{\sin(\theta)} \right) \,, \\
    \xi^{\mu}_{T_{z}} & = \left( 0, \chi\cos(\theta), -\frac{\chi}{r} \sin(\theta), 0 \right) \,, \\
    \xi^{\mu}_{R_{x}} & = \left( 0,0,\sin(\phi), \frac{\cos(\phi)}{\tan(\theta)} \right) \,, \\
    \xi^{\mu}_{R_{y}} & = \left( 0, 0, -\cos(\phi), \frac{\sin(\phi)}{\tan(\theta)} \right) \,, \\
    \xi^{\mu}_{R_{z}} & = \left(0, 0, 0, -1 \right) \,, 
\end{align}
\end{subequations}
where the parameter $\chi=\sqrt{1-kr^2}$ describes the curvature of the three-dimensional spatial surfaces. We restrict our analysis to the spatially flat case ($k=0$), implying  $\chi=1$. 
Imposing spatial homogeneity and isotropy in the metric can be regarded as imposing that the Lie derivative of the metric and the symmetric  teleparallel connection along the Killing vectors must vanish, that is 
\begin{equation}
\label{Liederivative}
\mathcal{L}_{\xi} g_{\mu\nu} = 0, \qquad \mathcal{L}_{\xi} \Gamma^{\rho}{}_{\mu\nu} = 0.
\end{equation}
The metric that satisfies this condition is the well-known Friedmann-Lemaître-Robertson-Walker metric, given by
\begin{align}
\label{metric}
    ds^2 &= -dt^2 + a(t)^2 \left(dr^2 + r^2 d\theta^2 + r^2 \sin^2 \theta d\phi \right) \,.
\end{align}
A matter energy momentum tensor consistent with this cosmological symmetry takes the form:
\begin{align}
\label{matter}
    \mathcal{T}_{\mu\nu} &= \left(\begin{matrix}\rho(t) & 0 & 0 & 0\\0 & a^{2}(t) p(t) & 0 & 0\\0 & 0 & r^{2} a^{2}(t) p(t) & 0\\0 & 0 & 0 & r^{2} a^{2}(t) p(t) \sin^{2}\theta \end{matrix}\right) \,,
\end{align}
where a barotropic equation of state $p = \mathrm{w} \rho$ is assumed.
The symmetric teleparallel connection components compatible with the spacetime symmetry \eqref{Killing_Vector} 
were worked out practically simultaneously in Refs.\ \cite{Hohmann:2021ast} and \cite{DAmbrosio:2021pnd}. In the following, we use the notation of Ref.\ \cite{Dimakis:2022rkd} that has become recurrent in the literature.

\subsection{Connection set 1} \label{Connection set 1}
We present the first set of spatially homogeneous and isotropic  symmetric teleparallel connection, which, under the assumption of zero spatial curvature (i.e., $k=0$) can be expressed as 

\begin{align}
\label{connectionset1}
    \Gamma^\rho{}_{\mu\nu} &= \left[\begin{matrix}\left[\begin{matrix}\gamma(t) & 0 & 0 & 0\\0 & 0 & 0 & 0\\0 & 0 & 0 & 0\\0 & 0 & 0 & 0\end{matrix}\right] & \left[\begin{matrix}0 & 0 & 0 & 0\\0 & 0 & 0 & 0\\0 & 0 & - r & 0\\0 & 0 & 0 & - r \sin^{2}\theta\end{matrix}\right] & \left[\begin{matrix}0 & 0 & 0 & 0\\0 & 0 & \frac{1}{r} & 0\\0 & \frac{1}{r} & 0 & 0\\0 & 0 & 0 & - \sin\theta \cos\theta\end{matrix}\right] & \left[\begin{matrix}0 & 0 & 0 & 0\\0 & 0 & 0 & \frac{1}{r}\\0 & 0 & 0 & \cot\theta\\0 & \frac{1}{r} & \cot\theta & 0\end{matrix}\right]\end{matrix}\right] \,,
\end{align}
where the four matrices in columns are labeled by the first index ${}^\rho$, and the entries of the matrices are specified by the last two indices ${}_{\mu \nu}$. This set was treated as case 1 ($K_2=K_3=0$) with $\gamma = K_1 = -K$ in Ref.\ \cite{Hohmann:2021ast}, and as case $\Gamma^{(III)}_Q$ with $\gamma=C_1$ in Ref.\  \cite{DAmbrosio:2021pnd}, while in Ref.\ \cite{Gomes:2023tur} it was dubbed the trivial branch (whereby $\alpha_{I}=\alpha_{II}=0$). There are no extra restrictions on the function $\gamma(t)$. The nonmetricity scalar \eqref{STEGRnonmetricity} computed from the connection \eqref{connectionset1} is 
\begin{align}\label{set1nonmetricity}
    Q &= -6 H^2\,.
\end{align}

Substituting the parametrization \eqref{eq:param}, metric \eqref{metric}, matter energy momentum tensor \eqref{matter} and connection \eqref{connectionset1} in the field equations \eqref{eq: fieldequationf(Q)}, we obtain the following set of cosmological equations 
\begin{subequations}
\label{eq: set1_F(Q)}
\begin{align}
\label{eq: FR1set1F(Q)}
    6 H^{2}+ 12 H^{2} F_{Q} + F = 2 \kappa^{2} \rho,  \\
    \label{eq: FR2set1F(Q)}
    - (4 \dot{H} + 6 H^{2})- 12 H^{2} F_{Q} + 48 H^{2}\dot{H}   F_{QQ} - 4 \dot{H} F_{Q}   - F = 2 \kappa^{2} \mathrm{w} \rho,\\
    \label{eq: Matterset1F(Q)}
    \dot{\rho}+3H(1+\mathrm{w})\rho = 0,
\end{align}
\end{subequations}
where subscripts denote derivatives, i.e.,
\begin{align}
    F_{Q} = \frac{dF}{dQ},\qquad F_{QQ} = \frac{d^{2}F}{dQ^{2}}.
\end{align}
The connection equation \eqref{connectionfieldeq} is identically satisfied by the expression \eqref{connectionset1} for any model $f(Q)$. Here the function $\gamma(t)$ is completely absent in the equations of motion and the equations \eqref{eq: set1_F(Q)} coincide precisely with the field equations derived assuming the coincident gauge (i.e.\ globally vanishing connection) in Cartesian coordinates \cite{BeltranJimenez:2017tkd}. Interestingly, as first pointed out in Ref.\ \cite{Jarv:2018bgs}, these equations also coincide with the metric teleparallel $f(T)$ gravity cosmological equations of the same symmetry 
\cite{Bengochea:2008gz,Linder:2010py,Hohmann:2019nat}. This means that for a given function $f$ the background evolution of $f(Q)$ and $f(T)$ models is the same, although the evolution of perturbations could differ.

Recall that before entering cosmology, we observed that when $Q$ is constant, the equations of motion \eqref{eq: fieldequationf(Q)} reduce to GR with a cosmological constant. It is worth mentioning that in the cosmology of connection set 1, the form of the nonmetricity scalar \eqref{set1nonmetricity} precludes reaching the GR limit in this manner (except for the Minkowski metric). If we want a constant or vanishing $Q$ that also exhibits the cosmological symmetry \eqref{Liederivative}, we need to take another set of connections. However, a trivial GR regime is still possible with the connection set 1 for model functions which allow $F(Q_*)=0$ for some value $Q_*$.

Since the function $\gamma$ drops out of the equations and the quantity $Q$ depends solely on $H$, the independent dynamics of the system \eqref{eq: set1_F(Q)} can be reduced to one dimension only, irrespective of the particular variables used \cite{Mirza:2017vrk,Hohmann:2017jao,Lu:2019hra,Boehmer:2023knj}. Although there are two first order differential equations, either the variable $H$ or $\rho$ can be expressed algebraically from the constraint \eqref{eq: FR1set1F(Q)}, reducing the number of independent dimensions of the phase space. Nevertheless, to compare with the alternative cosmological connection sets, we can write the system as
\begin{subequations}
\begin{align}
\label{eq: set1 H dot}
    \dot{H} =& \frac{-12 H^2 F_Q-F-6 H^2-2 \kappa ^2 \rho  w}{4 \left(-12 H^2 F_{\text{QQ}}+F_Q+1\right)} = - \frac{(1+\mathrm{w}) \kappa^2 \rho}{\frac{d}{dQ}(Q + 2Q F_Q - F)} \,,\\
    \dot{\rho} =& (1+\mathrm{w})\rho H.
\end{align}
\end{subequations}
Depending on the function $F$ the system may encounter a singularity where $\dot{H}$ diverges. Here the rhs of \eqref{eq: set1 H dot} is written in a way where the thorough analysis of singularities in Ref.\ \cite{Hohmann:2017jao} can be directly applied, as the systems in $f(Q)$ and $f(T)$ coincide.

\subsection{Connection set 2} \label{Connection set 2}
The second set of spatially homogeneous and isotropic symmetric teleparallel connection is given by
\begin{align}
\label{connectionset2}
   \Gamma^\rho{}_{\mu\nu} &= \left[\begin{matrix}\left[\begin{matrix}\gamma(t) + \frac{\dot{\gamma}(t)}{\gamma(t)} & 0 & 0 & 0\\0 & 0 & 0 & 0\\0 & 0 & 0 & 0\\0 & 0 & 0 & 0\end{matrix}\right] & \left[\begin{matrix}0 & \gamma(t) & 0 & 0\\\gamma(t) & 0 & 0 & 0\\0 & 0 & - r & 0\\0 & 0 & 0 & - r \sin^{2}\theta\end{matrix}\right] & \left[\begin{matrix}0 & 0 & \gamma(t) & 0\\0 & 0 & \frac{1}{r} & 0\\\gamma(t) & \frac{1}{r} & 0 & 0\\0 & 0 & 0 & - \sin\theta \cos\theta\end{matrix}\right] & \left[\begin{matrix}0 & 0 & 0 & \gamma(t)\\0 & 0 & 0 & \frac{1}{r}\\0 & 0 & 0 & \cot\theta\\\gamma(t) & \frac{1}{r} & \cot\theta & 0\end{matrix}\right]\end{matrix}\right] \,,
\end{align}
where by definition $\gamma(t) \neq 0$. This set was called case 3 ($K_2=0, K_3\neq 0$) with $\gamma = K_3 = K$ in Ref.\  \cite{Hohmann:2021ast} and case $\Gamma^{(I)}_Q$ with $\gamma=C_3$ in Ref.\ \cite{DAmbrosio:2021pnd}. The nonmetricity scalar \eqref{STEGRnonmetricity} characterizing the connection \eqref{connectionset2} is 
\begin{align}
\label{set2nonmetricity}
    Q &= -6 H^2 + 9 H \gamma + 3 \dot{\gamma}\,.
\end{align}
From here we can recognize the case $\alpha_{\text{I}}=1$, $\alpha_{\text{II}}=0$ for the nonmetricity scalar in Ref.\ \cite{Gomes:2023tur}. This implies that the nontrivial branch I of Ref.\ \cite{Gomes:2023tur} coincides with our connection set 2. Sometimes in the literature the case with the connection \eqref{connectionset2} is called as non-coincident gauge, in contrast to \eqref{connectionset1}. However, in spherical coordinates as presented here, both these connections are in a non-coincident gauge since the expression of $\Gamma^\rho{}_{\mu\nu}$ is not zero. By the procedure explained and exemplified in Ref.\ \cite{Bahamonde:2022zgj} it is possible to find a coordinate system  where the connection vanishes and the coincident gauge is realized. For the trivial connection set 1 \eqref{connectionset1} the coincident gauge is given by the simple Cartesian coordinates, while for the alternative connection set 2 \eqref{connectionset2} these coordinates are probably more involved and not yet reported in the literature.

Upon the substitution of the parametrization \eqref{eq:param}, metric \eqref{metric}, matter energy momentum tensor \eqref{matter} and connection \eqref{connectionset2} in the field equation \eqref{eq: fieldequationf(Q)}, we obtain the following set of cosmological equations:
    \begin{subequations}
        \label{connectionset2F(Q)}
            \begin{align}
        \label{eq: FR1set2F(Q)}
             6 H^{2} + F +  F_{Q} (6H^{2} -  Q) + 3 F_{QQ} \dot{Q} \gamma   =2 \kappa^{2} \rho \,,\\
         \label{eq: FR2set2F(Q)}
             - (6 H^{2} + 4 \dot{H}) - F + F_{Q}( - 6  H^{2} +  Q -  4  \dot{H}) + F_{QQ}\left(- 4  H \dot{Q} + 3 \dot{Q} \gamma \right)  =  2 \kappa^{2} \mathrm{w} \rho \,,\\
         \label{eq: Conn2F(Q)}
             3 \gamma \left(F_{QQQ} \dot{Q}^{2} + 3 F_{QQ} H \dot{Q} + F_{QQ} \ddot{Q}\right)= 0 \,,\\
        \label{eq: Matterset2F(Q)}
            \dot{\rho}+3H(1+\mathrm{w})\rho = 0 \,.
            \end{align}
    \end{subequations}
It is important to note the equations above are related to each other,
as mentioned before in Sec.\ \ref{Symmetric teleparallelism}. For instance, the Bianchi identity allows us to derive the connection equation by taking the time derivative of the first Friedmann equation \eqref{eq: FR1set2F(Q)} and using both the second Friedmann equation \eqref{eq: FR2set2F(Q)} and the matter continuity equation \eqref{eq: Matterset2F(Q)}. Therefore, the connection equation \eqref{eq: Conn2F(Q)} is not independent and does not introduce new dynamics, but is rather a constraint implied by the other equations.
However, for the sake of completeness, it is still useful to present it explicitly.

In contrast to connection set 1, the nonmetricity scalar \eqref{set2nonmetricity} can vanish or be constant for any $H(t)$ depending on the behavior of $\gamma(t)$. Then the equations \eqref{eq: FR1set2F(Q)}, \eqref{eq: FR2set2F(Q)} and \eqref{eq: Matterset2F(Q)} reduce to GR or GR+$\Lambda$ while the connection equation \eqref{eq: Conn2F(Q)} is identically satisfied. Again, a trivial GR regime is also possible with the connection set 2, for model functions which allow $F(Q_*)=0$ for some value $Q_*$.

The nonmetricity scalar \eqref{set2nonmetricity}  
encodes the dynamics of the extra function $\gamma$. Additionally, both Friedmann equations \eqref{eq: FR1set2F(Q)} and \eqref{eq: FR2set2F(Q)} contain a term involving the time derivative of the nonmetricity scalar, which contains the second derivative of $\gamma$. Therefore, it is of interest to determine the number of independent dynamical variables contained within the combined system of equations. We can rewrite the system \eqref{connectionset2F(Q)} as four first order differential equations
\begin{subequations}
\label{eq: set2 syn sys}
\begin{align}
\label{set2_gamma}
    \dot{\gamma }=&\Pi\,,\\
\label{set2_Pi}
    \dot{\Pi }= &\frac{F_Q \left(2 H-3 F_{\text{QQ}} \left(3 \gamma  \Pi +32 H^3-30 \gamma  H^2+9 \gamma ^2 H-4 H \Pi \right)\right)+F_{\text{QQ}} \Big(F (3 \gamma -6 H)-36 H^3+18 \gamma  H^2-6 H \Pi -4 H \kappa ^2 \rho  (\mathrm{w}-2)}{2 \left(F_Q+1\right) F_{\text{QQ}}}\nonumber\\
    &\frac{+3 \gamma  \kappa ^2 \rho  (\mathrm{w}-1)\Big)+2 H F_Q^2}{2 \left(F_Q+1\right) F_{\text{QQ}}}-\frac{\left(-12 H^2 F_{\text{QQ}}+F_Q+1\right) \left(3 F_Q \left(4 H^2-\Pi \right)+F+6 H^2-2 \kappa ^2 \rho \right)}{9 \gamma  \left(F_Q+1\right) F_{\text{QQ}}}\,, \\
\label{set2_H}
    \dot{H}=&\frac{2 F H -3 \gamma  F-12 H^3+36 \gamma  H^2-27 \gamma ^2 H -9 \gamma  \Pi +6 H \Pi +\rho  \left(3 \gamma  \kappa ^2-4 H \kappa ^2-3 \gamma  \kappa ^2 w\right)}{6 \gamma  \left(F_Q+1\right)}+\frac{(2 H-3 \gamma ) \left(4 H^2-3 \gamma  H-\Pi \right)}{2 \gamma }\,,\\
\label{set2_rho}
    \dot{\rho }= &-3 H \rho  (\mathrm{w}+1)\,.
\end{align}
\end{subequations}

Unlike the connection set 1 described in Sec.\ \ref{Connection set 1}, the nonmetricity scalar $Q$ in set 2 contains additional dynamics for $\gamma$ that prevents the Friedmann equation \eqref{eq: FR1set2F(Q)} from being a constraint equation. Hence, the total system can be described by four independent dynamical variables: ${\gamma}$, $\Pi$, $H$ and $\rho$, which span a four dimensional phase space. A further reduction might be achievable with specific choices of model, as detailed in \cite{Shabani:2023nvm,Paliathanasis:2023gfq}.

As we have seen in set 1, the singularity may arise depending on the function $F$; however, in set 2 the expression \eqref{set2_H} reveals that a singularity may happen either when $\gamma$ goes to zero or where the term $1+F_{Q}$ vanishes. The latter is in accord with the general discussion of the physicality condition around Eq.\ \eqref{eq: condition f_Q>0}. 
Similarly to the symmetric teleparallel scalar-tensor theories \cite{Jarv:2023sbp} a further warning about the possible singularities comes from the inspection of the connection equation \eqref{eq: Conn2F(Q)}. It can be equivalently written as
\begin{align}
\label{eq: nonmetricityfoce2}
    \ddot{Q}=-3H\dot{Q} - \frac{F_{QQQ}}{F_{QQ}}\dot{Q}^{2}
\end{align}
and analyzed in analogy to a simple mechanical system. The rhs of Eq.\ \eqref{eq: nonmetricityfoce2} represents a force that gives rise to an acceleration $\ddot{Q}$. In an expanding universe, the first term on the rhs dominates for small values of $H$. This term acts as a friction term, damping the expansion rate. 
For larger values of $\dot{Q}$, the second term dominates. The sign of $\frac{F_{QQQ}}{F_{QQ}}$ determines whether this term drives $\dot{Q}$ to larger or smaller values. Therefore, the vanishing of $F_{QQ}$ could signal a singularity, a point where the forces are infinite. 
Later in Sec.\ \ref{Numerical example}, studying a numerical example with the appropriate initial conditions, we will elucidate the occurrence of singularities in this system.

\subsection{Connection set 3} \label{Connection set 3}
The third set of spatially homogeneous and isotropic $k=0$ symmetric teleparallel connections can be presented as 
\begin{align}
\label{connectionset3}
    \Gamma^\rho{}_{\mu\nu} &= \left[\begin{matrix}\left[\begin{matrix}- \frac{\dot{\gamma}(t)}{\gamma(t)} & 0 & 0 & 0\\0 & \gamma(t) & 0 & 0\\0 & 0 & r^{2} \gamma(t) & 0\\0 & 0 & 0 & r^{2} \gamma(t) \sin^{2}\theta\end{matrix}\right] & \left[\begin{matrix}0 & 0 & 0 & 0\\0 & 0 & 0 & 0\\0 & 0 & - r & 0\\0 & 0 & 0 & - r \sin^{2}\theta\end{matrix}\right] & \left[\begin{matrix}0 & 0 & 0 & 0\\0 & 0 & \frac{1}{r} & 0\\0 & \frac{1}{r} & 0 & 0\\0 & 0 & 0 & - \sin\theta \cos\theta\end{matrix}\right] & \left[\begin{matrix}0 & 0 & 0 & 0\\0 & 0 & 0 & \frac{1}{r}\\0 & 0 & 0 & \cot\theta\\0 & \frac{1}{r} & \cot\theta & 0\end{matrix}\right]\end{matrix}\right],
\end{align}
where by definition $\gamma(t) \neq 0$. This set was dubbed case 2 ($K_2\neq0, K_3=0$) with $\gamma = K_2 = -a^2 K$ in Ref.\ \cite{Hohmann:2021ast} and case $\Gamma^{(II)}_Q$ with $\gamma=C_2$ in Ref.\ \cite{DAmbrosio:2021pnd}. Utilizing the connection \eqref{connectionset3}, we can obtain the corresponding  nonmetricity scalar \eqref{STEGRnonmetricity} as
\begin{align}
\label{set3nonmetricity}
Q &= -6 H^2 + 9 H \bar\gamma + 3 \dot{\bar\gamma},
\end{align}
where $\bar\gamma = \frac{\gamma(t)}{a(t)^2}$.
We find that in Ref.\ \cite{Gomes:2023tur} this connection corresponds to the nontrivial branch II with $\alpha_{\text{I}}=0$ and $\alpha_{\text{II}}=1$. 

Substituting the aforementioned parametrization \eqref{eq:param}, metric \eqref{metric}, matter energy momentum tensor \eqref{matter} and connection \eqref{connectionset3} in the field equations \eqref{eq: fieldequationf(Q)}, we arrive at the following set of cosmological equations
 \begin{subequations}
\label{connectionset3F(Q)}
    \begin{align}
    \label{eq: FR1set3F(Q)}
        6 H^{2} +F +  F_{Q}\left( 6H^{2} - Q\right) - 3 F_{QQ} \dot{Q} \bar{\gamma}  &= 2 \kappa^{2} \rho \,,\\
    \label{eq: FR2set3F(Q)}
      -( 6 H^{2} + 4 \dot{H} ) - F + F_{Q}\left( - 6  H^{2} +  Q - 4  \dot{H}\right) + F_{QQ} \left( - 4  H \dot{Q} + \bar{\gamma} \dot{Q}\right) &= 2 \kappa^{2} \mathrm{w} \rho \,,\\
\label{eq: Conn3F(Q)}
     - 3 \cdot \left(2 F_{QQ} \dot{Q} \dot{\bar{\gamma}} + \bar{\gamma} \left(F_{QQQ} \dot{Q}^{2} + 5 F_{QQ} H \dot{Q} + F_{QQ} \ddot{Q}\right)\right) &= 0  \,,\\
\label{eq: MC_Q2}
     \dot{\rho}+3H(1+\mathrm{w})\rho &= 0 \,.
\end{align}
\end{subequations}

Analogous to the connection set 2 in Sec.\ \ref{Connection set 2}, the connection equation \eqref{eq: Conn3F(Q)} depends on the remaining equations, implying that not all four equations \eqref{connectionset3F(Q)} are independent. Furthermore, due to the presence of the additional dynamical term $\dot{\bar{\gamma}}$ in the nonmetricity scalar \eqref{set3nonmetricity}, the Friedmann equation \eqref{eq: FR1set3F(Q)} is no longer a constraint equation. 

Just like in connection set 2, the nonmetricity scalar \eqref{set3nonmetricity} has a similar structure with the set 2 nonmetricity scalar \eqref{set2nonmetricity}. Hence, in cosmology when the nonmetricity scalar \eqref{set3nonmetricity} is vanishing or constant, the cosmological equation \eqref{eq: FR1set3F(Q)}-\eqref{eq: MC_Q2} successfully recovers GR or GR+$\Lambda$ and the connection equation \eqref{eq: Conn3F(Q)} is identically satisfied. A trivial regime which follows from $F(Q_{*})=0$ is also possible.

The total phase space can be represented by the following four dynamical variables:
    \begin{subequations}
        \label{Con3_Ind_Dyn}
            \begin{align}
        \label{Con3_gamma}
            \dot{\bar{\gamma} } = &\Pi\,,\\
        \label{Con3_pi}
            \dot{\Pi } = &\frac{6 F_Q^2 \left(4 H^2-3 \bar{\gamma}  H-\Pi \right)+F_Q \left(2 \left(F+18 H^2-9 \bar{\gamma}  H-2 \kappa ^2 \rho -3 \Pi \right)-9 H F_{\text{QQ}} \left(9 \bar{\gamma} ^3+8 \bar{\gamma}  \Pi +32 H^3-32 \bar{\gamma}  H^2-6 \bar{\gamma} ^2 H-8 H \Pi \right)\right)}{18 \bar{\gamma}  \left(F_Q+1\right) F_{\text{QQ}}}\nonumber\\
            &\frac{-3 H F_{\text{QQ}} \left(8 \bar{\gamma}  \kappa ^2 \rho +9 \bar{\gamma}  \Pi -\bar{\gamma}  F+4 F H+24 H^3-6\bar{\gamma}  H^2-9 \bar{\gamma} ^2 H-8 H \kappa ^2 \rho +6 \bar{\gamma}  \kappa ^2 \rho  \mathrm{w}\right)+F+6 H^2-2 \kappa ^2 \rho }{9 \bar{\gamma}  \left(F_Q+1\right) F_{\text{QQ}}}\,,\\
        \label{Con3_H}
            \dot{H} =&\frac{-3 \bar{\gamma}  \Pi -\bar{\gamma}  F-2 F H+12 H^3-12 \bar{\gamma}  H^2-9 \bar{\gamma} ^2 H-6 H \Pi +\rho  \left(-\bar{\gamma}  \kappa ^2+4 H \kappa ^2-3 \bar{\gamma}  \kappa ^2 w\right)}{6\bar{\gamma}  \left(F_Q+1\right)}+\frac{(\bar{\gamma} +2 H) \left(-4 H^2+3 \bar{\gamma} H+\Pi \right)}{2 \bar{\gamma} }\,,\\
            \dot{\rho }= &3 H \rho  (\mathrm{w}+1)\,.
            \end{align}
    \end{subequations}
Therefore, these four dynamical variables fully describe our four dimensional phase space. Further reduction might be achievable with specific choices of model functions, as detailed in \cite{Shabani:2023nvm,Paliathanasis:2023gfq}.

By looking at Eqs.\ \eqref{Con3_H}, a singularity could arise when either $\bar{\gamma}$ approaches zero or if the function $F_{Q}+1$ vanishes. These conditions likely lead to a divergence in the Hubble parameter, subsequently causing the whole system to hit a singularity.
The connection equation  \eqref{eq: Conn3F(Q)} can be equivalently expressed as
    \begin{align}
        \label{eq: nonmetricityfoce3}
            \ddot{Q} = -\left(\frac{2 \dot{\bar{\gamma}}}{\bar{\gamma} }+5 H\right)\dot{Q} - \frac{ F_{\text{QQQ}}}{F_{\text{QQ}}}\dot{Q}^2.
    \end{align}
Equation \eqref{eq: nonmetricityfoce3} shares the same structure as equation \eqref{eq: nonmetricityfoce2}  in set 2. The only difference is the addition of a friction term in equation \eqref{eq: nonmetricityfoce3} that depends on the parameter $\bar{\gamma}$. As a result, all the possible singularities discussed for set 2 are also relevant for the set 3.

\section{General Relativity Limit \label{General Relativity Limit}}
In this section, we investigate the conditions under which the modified symmetric teleparallel field equations \eqref{eq: FR1set1F(Q)}, \eqref{connectionset2F(Q)} and \eqref{connectionset3F(Q)} for the FLRW metric reduce to general relativity.

\subsection{Relaxation to general relativity \label{Relaxation to general relativity}}
The well-known Friedmann equations for a single barotropic fluid $(p = \mathrm{w} \rho)$ along with the matter continuity equation in general relativity are represented by 
    \begin{subequations}
        \label{eq: GR}
            \begin{align}
        \label{eq: FR1_GR}
            3 H{}_{*}(t)^2   &=  \kappa ^2 \rho_{*}(t)  \,,\\
        \label{eq: FR2_GR}
            -(3 H{}_{*}(t)^2+2 \dot{H}_{*}(t))  &=  \kappa ^2 \mathrm{w}  \rho _{*}(t)\,,\\
        \label{eq: MC_GR}
            \dot{\rho}_{*}(t) + 3H_{*}(t)\rho_{*}(t) (1+\mathrm{w}) &=0\,.
            \end{align}
    \end{subequations}
We explore the possible scenario under which the symmetric teleparallel metric field equations recover general relativity. Comparing our metric equations \eqref{eq: FR1set1F(Q)}--\eqref{eq: FR2set1F(Q)} of set 1, \eqref{eq: FR1set2F(Q)}--\eqref{eq: FR2set2F(Q)} of set 2 and \eqref{eq: FR1set3F(Q)}--\eqref{eq: FR2set3F(Q)} of set 3 with the GR equations \eqref{eq: FR1_GR}--\eqref{eq: FR2_GR}, we find some additional terms, which we will refer as the non general relativity (nGR) part. When $F$ goes to zero or the nGR contribution vanishes, our metric equations reduce to the GR equations \eqref{eq: FR1_GR}--\eqref{eq: FR2_GR}, which is what we will call the GR limit. Notably,  in the latter case, a correction is introduced to the full field equations, potentially leading to significant consequences for the evolution of perturbations within the cosmological background. We will explore these consequences in the upcoming Sec.\ \ref{Connection set I}.

According to \eqref{eq: GR}, for nonrelativistic or dust matter dominated cases ($\mathrm{w}=0$) the Hubble parameter and matter energy density evolve as 
    \begin{align}
        \label{eq: bgmatter}
             H_{*}(t)=\frac{2}{3(t-t_{0})},\quad \quad \rho_{*}(t)=\frac{4}{3\kappa^{2}(t-t_{0})^{2}} \,.
    \end{align}
For relativistic or radiation dominated case $\left(\mathrm{w}=\frac{1}{3}\right)$ we have
\begin{align}
\label{eq: bgradiation}
H_{*}(t)=\frac{1}{2(t-t_{0})},\quad \quad \rho_{*}(t)=\frac{3}{4\kappa^{2}(t-t_{0})^2} \,.  
\end{align}

To constrain the nature of dark energy, we employ two distinct theoretical frameworks in our analysis. In the first approach, dark energy is investigated by assuming constant energy density ($\rho_{*}=\text{constant}= \Lambda_{DE}$) with an equation of state $\mathrm{w} = -1$, known as dark energy dominated era. The Hubble parameter and energy density become 
    \begin{align}
        \label{eq: bgdarkenergy}
            H_{*}(t)=\pm\sqrt{\frac{\kappa^{2}\Lambda_{DE}}{3}},\quad \quad \rho_{*}(t) = \Lambda_{DE}.
    \end{align} 
In the second approach, we explore the case where the dark energy is purely geometric and dominates over any matter component, thus $\rho_{*}=0$. Unlike other epochs, the general relativity  equations \eqref{eq: GR} incorporate a cosmological constant term, $\Lambda_{GDE}$, and the standard GR equations adopt the following form as 
    \begin{subequations}
        \label{eq: bg_gddark}
            \begin{align}
        \label{eq: FR1_gddark}
            3 H{}_{*}(t)^2 - \kappa ^{2}\Lambda_{GDE}   &=  0  \,,\\
        \label{eq: FR2_gddark}
            -(3 H{}_{*}(t)^2+2 \dot{H}_{*}(t)) - \kappa ^{2} \Lambda_{GDE}  &=  0\,,
            \end{align}
    \end{subequations}
while the matter continuity equation remains same as \eqref{eq: GR}. Hence, the Hubble function after solving \eqref{eq: bg_gddark} yields
    \begin{align}
        \label{eq: bggeomdarkenergy}
            H_{*}(t) = \pm\sqrt{\frac{\kappa^{2}\Lambda_{GDE} }{3}}, \qquad \rho_{*}(t)= 0.
    \end{align} 
In this context, $H_{*}(t)$ and $\rho_{*}(t)$ evolve in time and $t_{0}$ as an integration constant that sets the initial reference point for the timescale. For convenience, in the following analysis we set $t_{0}=0$. As usual, the time variable $t$ is assumed to be monotonically increasing ($t>0$). Since the density of matter types $\rho_{*}(t)$ evolves at three different rates, our Universe goes through the sequence of radiation, dust matter, and dark energy dominated eras where other contributions are considered as much less significant.

To investigate the dynamical stability of the universe across its evolutionary eras, we employ an approximation scheme near the general relativity limit through the following expansions
    \begin{align}
        \label{solexpansion}
            Q(t)=Q_{*}(t)+q(t) \,, \qquad H(t)= H_{*}(t)+h(t) \,, \qquad \gamma(t)=\gamma_{*}(t)+g(t) \,, \qquad \rho(t)=\rho_{*}(t)+r(t)\,,
    \end{align}
where $q(t)$, $h(t)$, $r(t)$ and $g(t)$ are small perturbations assumed to be of approximately the same order.
Again through a Taylor series expansion, we approximate the function $F(Q)$ depending on the nonmetricity scalar around the background value of the nonmetricity scalar $Q_{*}(t)$ as follows 
    \begin{subequations}
        \label{functionalexpansion}
            \begin{align}
        \label{function1expansion1}
             F(Q(t)) &=F^*(t)+F^{*}_Q(t)q(t)+\frac{1}{2} F^{*}_{\text{QQ}}(t) q(t)^2 \,, \\
         \label{function1expansion2}
             F_{Q}(Q(t)) &=F^{*}_Q(t)+F^{*}_{\text{QQ}}(t)q(t)+\frac{1}{2}F^{*}_{\text{QQQ}}(t) q(t)^2 \,, \\
        \label{function1expansion3}
            F_{QQ}(Q(t)) &=F^{*}_{\text{QQ}}(t)+F^{*}_{\text{QQQ}}(t) q(t)+\frac{1}{2}F^{*}_{\text{QQQQ}}(t) q(t)^2\,,  \\
        \label{function1expansion4}
            F_{QQQ}(Q(t)) &=F^{*}_{\text{QQQ}}(t)+F^{*}_{\text{QQQQ}}(t) q(t)+\frac{1}{2}F^{*}_{\text{QQQQQ}}(t) q(t)^2\,,
            \end{align}
    \end{subequations}
where the superscript $^{*}$ denotes the value computed at the background value $Q_{*}(t)$. 

This work investigates the possibility of a stable scenario around the GR limit, i.e.\ a safe realization of radiation, dust matter, and dark energy dominated eras. In principle, it would be possible to include extra fluids which are subdominant and could be considered as an additional perturbation. However, the inclusion of the extra fluid could naturally lead to a transition between two eras which is a more involved process. Only if our gravitational theory can accommodate the well-established cosmological eras in a stable manner, an investigation of the transition between the eras will become relevant and would then constitute a separate task.

We begin by substituting expansions \eqref{solexpansion} and approximate function \eqref{functionalexpansion}  into the cosmological equations \eqref{eq: set1_F(Q)}, \eqref{connectionset2F(Q)} and \eqref{connectionset3F(Q)} to get an approximate field equations, which can be solved analytically order by order. At the background order for all three sets, the metric and matter field equations reduce to \eqref{eq: GR}, and the connection equations are identically satisfied. 
To analyze the behavior of the perturbed function from the radiation to the dark energy dominated era in both connection sets 1 and 2, we utilize \eqref{solexpansion}, functional expansion of nonmetricity \eqref{function1expansion1}-\eqref{function1expansion4} and substitute it into the cosmological equations \eqref{eq: set1_F(Q)} and \eqref{connectionset2F(Q)}. Following this, we will solve the expanded equations systematically in a perturbative manner, treating each order of magnitude separately. We begin by investigating the background evolution, which means all perturbed functions are negligible. Notably, while the equations of motion of $f(Q)$ in set 1 resembles $f(T)$, our analysis is new for connection set 2.

\section{Connection set 1 in the general relativity limit \label{Connection set I}}
In this section we investigate a suitable $F(Q)$ function compatible with a GR background, and we analyze whether the perturbations have a stable behavior around the GR solution in all the epochs of the  standard history of our Universe. 

\subsection{Background evolution}
\label{Background I}
Using the method introduced in \eqref{Relaxation to general relativity}, 
we first compare the set 1 equation \eqref{eq: FR1set1F(Q)} to the corresponding equation \eqref{eq: FR1_GR} of general relativity, and identify the nGR term. Together, the nGR parts are
    \begin{subequations}
        \begin{align}
    \label{eq: bgenergy_set1}
        12 H_*(t){}^2 F^{*}_Q{}(t)+F^*(t)=0\,,\\
    \label{eq: bgpressure_set1}
        -F^*(t)+\left(-12 H_*(t){}^2-4 \dot{H}_*(t)\right) F^*_Q(t)+48 \dot{H}_*(t) H_*(t){}^2 F^*_{QQ}(t) =0\,.
    \end{align}
\end{subequations}
When $F^*(t)$ is zero, the nGR term identically vanishes, henceforth Eq.\ \eqref{eq: FR1set1F(Q)} becomes the standard Friedmann equations of STEGR and GR. For a nonzero $F^*(t)$, we solve the nGR equation, which leads to a nontrivial solution
\begin{align}
\label{eq: set1 F}
    F^*(t) &=\alpha\sqrt{-Q_{*}(t)}
\end{align} 
where $\alpha$ is a constant, and the field equations \eqref{eq: FR1set1F(Q)} still become  identical to STEGR and GR. Finding this form is not entirely surprising, as it was already noted in the $f(T)$ context that with this model function the FLRW equations reduce to GR \cite{Hohmann:2017jao}.

Our analysis has so far yielded the following known background quantities: the Hubble expansion $H_{*}$, energy density $\rho_{*}$ and the background value of the nonmetricity scalar $Q_{*}=-6H_{*}(t)^{2}$. Using Eq.\ \eqref{eq: bgenergy_set1} we can write the first and second derivative of $F^{*}(t)$ with respect to $Q_{*}$ as
    \begin{align}
        \label{eq: F derivatives set 1}
             F^{*}_{Q}(t)=\frac{F^{*}(t)}{2Q_{*}(t)}\,,\qquad F^{*}_{QQ}(t)=-\frac{ F^{*}(t)}{4Q_{*}(t)^{2}} \,,
    \end{align}
which plays an important role in solving the approximate field equations. As it can be seen, Eq.\ \eqref{eq: bgenergy_set1} is easily solvable for the function of nonmetricity irrespective of any specific Hubble parameter or matter content. This suggests that the same function is applicable across different cosmological eras, i.e.\ models which suitably approximate \eqref{eq: set1 F} have a good chance in describing well the whole history of the Universe. 

For the hypothetic geometric dark energy era, we adopt the same strategy to identify the nGR background evolution. We achieve this by comparing the modified $F(Q)$ equation \eqref{eq: set1_F(Q)} with the standard Friedmann equation of geometric dark energy \eqref{eq: bg_gddark}.
As a result, the effective cosmological constant term can be written as
    \begin{align}
        \label{eq: effective_gde}
            \Lambda_{GDE} = \frac{-\Big(12 H_*(t){}^2 F^{*}_Q(t)+F^*(t)\Big)}{2\kappa^{2}}.
    \end{align}
The above equation \eqref{eq: effective_gde} allows us to write the subsequent derivative term as,
    \begin{align}
        \label{eq: gde_F derivatives set 1}
            F^{*}_{Q}(t)= \frac{F^*(t)+2 \Lambda \kappa^{2}}{2 Q_*(t)} \,,\qquad F^{*}_{QQ}(t)= \frac{Q_*(t) \left(F_Q\right){}^*(t) - \big(F^*(t) + 2 \Lambda \kappa^{2}\big)}{2 Q_*(t){}^2},
    \end{align}
which will be later substituted in the field equation to evaluate the perturbed functions.

Up to this point, we have successfully determined all the background quantities. To assess the behavior of the small perturbations $h(t)$, $r(t)$, $g(t)$, and $q(t)$ we must delve into the perturbed equations at first order. Our objective is to examine the convergence of these small perturbations over a long run of time, which sums up to determine whether a stable scenario is possible around the GR limit of extended symmetric teleparallel gravity, a study that we elaborated upon in the subsequent subsection.

\subsection{Stability of the standard cosmological regimes} \label{Stability of the standard cosmological regimes}
Having established the theoretical framework, we are now ready to examine the possibility of a stable transition from the radiation to the dark energy era.
First, we aim to eliminate the small perturbation of the nonmetricity scalar denoted by $q(t)$, which is not an independent quantity. This occurs because of the relationship between the background nonmetricity scalar, $Q_{*}(t)=-6H_{*}(t)^{2}$ and the general nonmetricity scalar $Q(t)=-6H(t)^{2}$, via the nonmetricity expansion 
\begin{align}
   Q(t) &=Q_{*}(t) + q(t) \,.  
\end{align}
This allows us to express the dependent quantity $q(t)$ in terms of the Hubble perturbation $h(t)$ and the background Hubble function, which is
\begin{align}
\label{expansion of small q}
    q(t)=-6 \Big( 2H_{*}(t)h(t)+h(t)^{2} \Big).
\end{align}

By substituting \eqref{solexpansion}, \eqref{expansion of small q}, and \eqref{function1expansion1}-\eqref{function1expansion4} into the connection set 1 cosmological equations \eqref{eq: FR1set1F(Q)}-\eqref{eq: Matterset1F(Q)},
we obtain up to first-order small quantities
\begin{subequations}
\label{eq: set1expansion}
\begin{align}
    \label{eq: expandedFR1set1}
     12 h(t) H_*(t)-2 \kappa ^2 r(t) = 0\,,\\
     \label{eq: expandedFR2set1}
    -4 \dot{h}(t)-12 h(t) H_*(t)-2 \kappa ^2 \mathrm{w}  r(t) = 0\,,\\
     \label{eq: mattdensset1}
    3(1+\mathrm{w}) h(t) \rho _*(t)+3(1+\mathrm{w})  H_*(t) r(t)+\dot{r}(t) =0\,.
\end{align}
\end{subequations}
Hence taking into account the background value \eqref{eq: bgmatter} within \eqref{eq: expandedFR1set1}-\eqref{eq: mattdensset1} we can obtain the first order small perturbations as 
\begin{align}
    h(t) \sim \frac{c_{1}}{t^{2}},  \qquad r(t) \sim \frac{4c_{1}}{t^{3}\kappa^{2}}\,,
\end{align}
where the $c_i$'s here and below denote integration constants, which we keep explicitly to facilitate easy comparisons with the cases in the later sections.
Thus, the perturbations of the Hubble parameter and the matter density decrease in time, suggesting the possibility of a stable dust dominated regime.

Similar to the matter dominated era, we can substitute the corresponding background value for the radiation dominated era  \eqref{eq: bgradiation} into \eqref{eq: expandedFR1set1}-\eqref{eq: mattdensset1}. This yields the first order small perturbations: 
\begin{align}
    h(t) \sim \frac{c_{2}}{t^{2}}, \qquad   r(t) \sim \frac{3 c_{2}}{t^{3}\kappa^{2}}. 
\end{align}
This solution suggests an asymptotic decay of the perturbed quantities over time, implying the possibility of a stable radiation dominated era consistent with the general relativity solution.

Analogously to the previous cases, the perturbed solution for the dark energy dominated era can be obtained by
substituting the corresponding background evolution \eqref{eq: bgdarkenergy} into \eqref{eq: expandedFR1set1}-\eqref{eq: mattdensset1}. This yields the first order perturbed solution:
    \begin{align}
        h(t) \sim c_{3}, \qquad r(t) \sim c_{4}.
    \end{align}
Here, we observe that the perturbed quantities, $h(t)$ and $r(t)$, neither decay nor grow asymptotically with time. This suggests that a standard dark energy dominated era might lead to a marginally stable solution around the general relativity background.

In the geometric dark energy dominated era, we analyze the perturbed evolution of the universe by substituting the corresponding background solution \eqref{eq: bggeomdarkenergy} into \eqref{eq: expandedFR1set1}-\eqref{eq: mattdensset1}, which yields the following perturbed solution:
    \begin{align}
        h(t) \sim c_{5}e^{-3H_{*}t}, \qquad  r(t) \sim c_{6}e^{-3H_{*}t}.
    \end{align}
The observed exponential decay of both $h(t)$ and $r(t)$ over time suggests a cosmological scenario dominated by stable geometric dark energy, asymptotically approaching a general relativity regime. At this point one might ask why a significant difference is observed in the equations for the perturbations, although the background behavior of dark energy and geometric dark energy is the same. The reason is that in the GDE case there is no matter component of dark energy and the respective perturbations are included in the gravitational (geometric) sector only. Thus, to involve some perturbations in the matter sector we take a dust matter contribution. However, in a DE dominated era, the cosmological constant (vacuum energy) was taken as a component in the matter sector and in our scheme its perturbations are relevant. The same procedure is applied in the perturbed equations of connection set 2 as well.

Our analysis demonstrates that the same function $F^{*}(t)$ obtained from \eqref{eq: F derivatives set 1} captures stable radiation, matter, and geometric dark energy eras, transitioning to a marginally stable dark energy era. Notably, for the geometric dark energy era, the additional constant term with the function $F^{*}(t)$ \eqref{eq: gde_F derivatives set 1} does not alter the function behavior. We can easily verify, that for the trivial background, i.e., $F^{*}(t)=0$, and the corresponding perturbed equations up to linear order remain the same as \eqref{eq: set1expansion}. Hence, the corresponding stability analysis remains unaltered. 

In summary, the symmetric teleparallel models can predict a history of the universe with small deviations from GR, if the model function $F(Q)$ can enter a regime where it approximates the square root expression \eqref{eq: set1 F} or trivial expression $F^*(t)=0$. This can explain why so many different models can fit the observational data well \cite{Lazkoz:2019sjl,Ayuso:2020dcu,Anagnostopoulos:2021ydo,Atayde:2021pgb,Capozziello:2022wgl,Ferreira:2023awf,Shi:2023kvu,Yang:2024tkw,Wang:2024eai}, as they accommodate solutions which are just stable perturbations around the basic $\Lambda$CDM trajectory.

\section{Connection set 2 in the general relativity limit} \label{Connection set 2 in the general relativity limit}
\label{Connection set II}
The appearance of the additional free function $\gamma$ in the cosmological equation \eqref{connectionset2F(Q)} motivates us to investigate its influence on the background dynamics as well as whether it suggests any signs of background instability. 

\subsection{Background evolution}
\label{Background Evolution}
By substituting the background expansion \eqref{solexpansion} and functional expansion \eqref{functionalexpansion} into the symmetric teleparallel cosmological equations \eqref{connectionset2F(Q)}, the GR terms of the metric \eqref{eq: FR1set2F(Q)}, \eqref{eq: FR2set2F(Q)} and the matter equation \eqref{eq: Matterset2F(Q)}  are completely solved by using the background cosmological solution \eqref{eq: bgmatter}, \eqref{eq: bgradiation}, \eqref{eq: bgdarkenergy}, \eqref{eq: bgmatter} and \eqref{eq: bgdarkenergy}.
The remaining nGR terms, together with the connection equation, can be given as
\begin{subequations}
\label{eq: Set2_Bg_Expansion} 
\begin{align}
\label{eq: Set2_Bg_Expa_ED}
    F^*(t)+\Big(6 H_*(t){}^2-Q_*(t)\Big) F^{*}_Q(t)+3 \dot{Q}_*(t) \gamma _*(t) F^{*}_{QQ}(t) &=0 \,,\\
\label{eq: Set2_Bg_Expa_PR}
    -F^*(t)+F^{*}_{QQ}(t) \Big(3 \dot{Q}_*(t) \gamma _*(t)-4 H_*(t) \dot{Q}_*(t)\Big)+  \Big(-6 H_*(t){}^2-4 \dot{H}_*(t)+Q_*(t)\Big) F^{*}_Q(t) &=0 \,,\\
\label{eq: Set2_Bg_Expa_Conn}
    3 \Big(\gamma _*(t) F^{*}_{QQ}(t) \Big(3 H_*(t) \dot{Q}_*(t)+\ddot{Q}_*(t)\Big)+ \dot{Q}_*(t){}^2 \gamma _*(t) F^{*}_{QQQ}(t)\Big) &=0 \,.
\end{align}
\end{subequations}
In the trivial case when $F_{*}(t)=0$, the nGR equations are identically satisfied, and hence we recover GR which refers to the trivial GR limit. Unlike for the connection set 1, in the set 2 case we may expect that the other trivial GR limit arising from a constant or vanishing nonmetricity scalar is compatible with evolving $H(t)$. However, imposing such scenario into the nGR equations \eqref{eq: Set2_Bg_Expansion} implies that $H_*$ is constant (or zero), and thus only geometric dark energy with de Sitter expansion can be realized in this way.
It is more interesting to search for a nontrivial function $F^{*}(t)$ that could also completely solve the above nGR equations \eqref{eq: Set2_Bg_Expansion} and reduce to the general relativity solution. However, the presence of the unknown function $\gamma_{*}$ and its derivative in the Friedmann equation hinders a direct solution. Therefore, we will instead begin by solving the connection equation \eqref{eq: Set2_Bg_Expa_Conn}, which can be rewritten as
\begin{align}
\label{eq: bgconnection}
3 \gamma_{*}(t) \Big(F^{*}_{QQQ}(t) \dot{Q_{*}}(t)^{2} + (3 H_{*}(t) \dot{Q}_{*}(t) +  \ddot{Q}_{*}(t))F^{*}_{QQ}(t)\Big)= 0,
\end{align}
while we know $\gamma_{*}(t)$ is non zero by construction. Hence, the second term within the bracket must vanish identically.  For clarity, let us explicitly express the time derivative of the function of nonmetricity
\begin{align}
\label{eq: derivativeofbgfunction}
   \frac{d}{dt}F^{*}_{Q}(t) = F^{*}_{QQ}(t) \dot{Q}_{*}(t)\,, \qquad \frac{d}{dt}F^{*}_{QQ}(t) = F^{*}_{QQQ}(t)\dot{Q}_{*}(t)\,, \qquad  \frac{d}{dt}\left( F^{*}_{QQ}(t)\dot{Q}_{*}(t)\right) = F^{*}_{QQQ}(t) \dot{Q_{*}}(t)^{2} + F^{*}_{QQ}(t) \ddot{Q}_{*}(t),
\end{align}
which are useful when solving the connection equation \eqref{eq: bgconnection}. Now, substituting \eqref{eq: derivativeofbgfunction} into \eqref{eq: bgconnection} yields the following expression
\begin{align}
\label{Redef_Conn}
    \frac{d}{dt}\left( F^{*}_{QQ}(t)\dot{Q}_{*}(t)\right) + 3H_{*}(t)F^{*}_{QQ}(t) \dot{Q}_{*}(t) = 0.
\end{align}
After performing an integration over time, we get
\begin{align} 
\label{eq: intigratingconnection}
    \ln\left(F^{*}_{QQ}(t) \dot{Q}_{*}(t)\right) = -3\int H_{*}(t)dt.
\end{align}
Unlike the connection set 1, Eq.\ \eqref{eq: Set2_Bg_Expa_ED} and \eqref{eq: Set2_Bg_Expa_PR} involve $\gamma_{*}(t)$ and its time derivatives up to $\ddot{\gamma}_{*}(t)$, which leads to more complex expressions to extract any background evolution. Nevertheless, Eq.\ \eqref{eq: Set2_Bg_Expa_Conn} offers a simplified form suitable for solving the function of nonmetricity. To emphasize, Eq.\ \eqref{eq: intigratingconnection}, due to the dependence on the different background expansion history \eqref{eq: bgmatter}, \eqref{eq: bgradiation}, \eqref{eq: bgdarkenergy} and \eqref{eq: bggeomdarkenergy}, the function is expected to vary across each cosmological epoch. In contrast to the connection set 1, a single nonmetricity function cannot capture the entire evolutionary history of the universe, from radiation to dark energy domination, and the respective forms of $F^*(Q)$ are given
in Sec. \ref{Sec_Set2_BG_Matter}, \ref{Sec_Set2_BG_Rad}, \ref{Sec_Set2_BG_DE} and \ref{Sec_Set2_BG_GDE}.

\subsection{Perturbative expansion}
\label{Perturbed Expansion}
We are familiar with the background values of Hubble expansion and matter energy density, which are \eqref{eq: bgmatter}, \eqref{eq: bgradiation}, \eqref{eq: bgdarkenergy} and \eqref{eq: bggeomdarkenergy} throughout all the evolutionary eras. However, the distinguishing factor lies in the background value of $q(t)$ due to the altered nonmetricity scalar as expressed in Eq.\eqref{set2nonmetricity}. Straightforward substitution from known expansions \eqref{solexpansion} and background nonmetricity scalar
\begin{align}
    Q_{*}(t) = -6 H_{*}(t)^2 + 9 H_{*}(t)\gamma_{*}(t) + 3 \dot{\gamma}_{*}(t)
\end{align}
into nonmetricity scalar \eqref{set2nonmetricity} yields the perturbed function as
\begin{align} 
\label{eq: set2smallq(t)}
    q(t) = 3 \Big(3 g(t) H_*(t)+\dot{g}(t)-4 h(t) H_*(t)+3 h(t) \gamma _*(t)\Big)+3   \Big(3 g(t) h(t)-2 h(t)^2\Big).
\end{align}

Substituting the expansions \eqref{solexpansion} around the GR limit, functional expansions \eqref{functionalexpansion}, and perturbed function $q(t)$ \eqref{eq: set2smallq(t)} in the cosmological equations of set 2 \eqref{eq: FR1set2F(Q)}-\eqref{eq: Matterset2F(Q)}, and considering up to linear perturbed terms, we obtain the following set of cosmological equations:
\begin{subequations}
\label{eq: Set2_Pert_Expansion}
\begin{align}
\label{eq: Set2_Pert_Expa_ED}
    F^{*}_{QQ}(t) \Bigg(3 \Big(3 \gamma _*(t) \big(\ddot{g}(t)+\dot{h}(t) \left(3 \gamma _*(t)-4 H_*(t)\big)\right)+\dot{g}(t) \left(9 H_*(t) \gamma _*(t)+6 H_*(t){}^2-Q_*(t)\right)+g(t) \big(-3 H_*(t) Q_*(t) & \nonumber\\
     +9 \dot{H}_*(t) \gamma _*(t)+18 H_*(t){}^3+\dot{Q}_*(t)\big)\Big)-9 h(t) \gamma _*(t) \left(4 \dot{H}_*(t)+Q_*(t)-3 \dot{\gamma }_*(t)\right)+12 h(t) H_*(t) Q_*(t)+54 h(t) H_*(t){}^2 \gamma _*(t) & \nonumber\\
    -72 h(t) H_*(t){}^3\Bigg) + F^{*}_{QQQ}(t) \Bigg(9 \dot{Q}_*(t) \gamma _*(t) \left(3 g(t) H_*(t)+\dot{g}(t)\right)-36 h(t) H_*(t) \dot{Q}_*(t) \gamma _*(t) +27 h(t) \dot{Q}_*(t) \gamma _*(t){}^2\Bigg) & \nonumber\\
   +12 h(t) H_*(t) F^{*}_Q(t)+12 h(t) H_*(t)-2 \kappa ^2 r(t) &=0   \,,\\
\label{eq: Set2_Pert_Expa_PR}
    F^{*}_{QQ}(t) \Bigg(3 \Big(\dot{g}(t) \left(9 H_*(t) \left(\gamma _*(t)-2 H_*(t)\right)-4 \dot{H}_*(t)+Q_*(t)\right)+g(t) \big(3 H_*(t) \big(Q_*(t)-8 \dot{H}_*(t)\big)+9 \dot{H}_*(t) \gamma _*(t) & \nonumber\\
     -18 H_*(t){}^3+\dot{Q}_*(t)\big)+\ddot{g}(t) \left(3 \gamma _*(t)-4 H_*(t)\right)\Big)-12 h(t) H_*(t) \big(-8 \dot{H}_*(t)+Q_*(t)+3 \dot{\gamma }_*(t)\big) & \nonumber\\
     +h(t) \left(9 \gamma _*(t) \left(-8 \dot{H}_*(t)+Q_*(t)+3 \dot{\gamma }_*(t)\right)-4 \dot{Q}_*(t)\right)-54 h(t) H_*(t){}^2 \gamma _*(t)+3 \dot{h}(t) \big(4 H_*(t) & \nonumber\\
    -3 \gamma _*(t)\big){}^2+72 h(t) H_*(t){}^3\Bigg)+F^{*}_{QQQ}(t) \Bigg(-3 \dot{Q}_*(t) \left(3 g(t) H_*(t)+\dot{g}(t)\right) \left(4 H_*(t)-3 \gamma _*(t)\right)& \nonumber\\
    -72 h(t) H_*(t) \dot{Q}_*(t) \gamma _*(t)+48 h(t) H_*(t){}^2 \dot{Q}_*(t)+27 h(t) \dot{Q}_*(t) \gamma _*(t){}^2\Bigg)+\Big(-12 h(t) H_*(t)& \nonumber\\
    -4 \dot{h}(t)\Big) F^{*}_Q(t) -12 h(t) H_*(t)-4 \dot{h}(t)-2 \kappa ^2 \mathrm{w}  r(t)&=0 \,,\\
\label{eq: Set2_Pert_Expa_Conn}
     3 F^{*}_{QQ}(t) \Bigg(3 g(t) H_*(t) \dot{Q}_*(t)+18 H_*(t) \gamma _*(t) \ddot{g}(t)+9 g(t) \gamma _*(t) \ddot{H}_*(t)+27 \dot{g}(t) H_*(t){}^2 \gamma _*(t)+27 g(t) \dot{H}_*(t) H_*(t) \gamma _*(t)  & \nonumber \\
    +18 \dot{g}(t) \dot{H}_*(t) \gamma _*(t)+g(t) \ddot{Q}_*(t)+3 \gamma _*(t) \dddot{g}(t)-12 H_*(t) \gamma _*(t) \ddot{h}(t)-12 h(t) \gamma _*(t) \ddot{H}_*(t)-36 \dot{h}(t) H_*(t){}^2 \gamma _*(t)& \nonumber \\
    +27 \dot{h}(t) H_*(t) \gamma _*(t){}^2-36 h(t) \dot{H}_*(t) H_*(t) \gamma _*(t)+27 h(t) H_*(t) \gamma _*(t) \dot{\gamma }_*(t)-24 \dot{h}(t) \dot{H}_*(t) \gamma _*(t)+3 h(t) \dot{Q}_*(t) \gamma _*(t)  & \nonumber \\
      +9 \gamma _*(t){}^2 \ddot{h}(t)+9 h(t) \gamma _*(t) \gamma _*''(t)+18 \dot{h}(t) \gamma _*(t) \dot{\gamma }_*(t)\Bigg)+3 F^{*}_{QQQ}(t) \Bigg(9 g(t) H_*(t) \gamma _*(t) \ddot{Q}_*(t)+27 g(t) H_*(t){}^2 \dot{Q}_*(t) \gamma _*(t)& \nonumber \\
     +27 \dot{g}(t) H_*(t) \dot{Q}_*(t) \gamma _*(t)+18 g(t) \dot{H}_*(t) \dot{Q}_*(t) \gamma _*(t)+6 \dot{Q}_*(t) \gamma _*(t) \ddot{g}(t)+3 \dot{g}(t) \gamma _*(t) \ddot{Q}_*(t) +g(t) \dot{Q}_*(t){}^2& \nonumber \\
    -12 h(t) H_*(t) \gamma _*(t) \ddot{Q}_*(t)-36 h(t) H_*(t){}^2 \dot{Q}_*(t) \gamma _*(t)+27 h(t) H_*(t) \dot{Q}_*(t) \gamma _*(t){}^2& \nonumber \\
     -24 \dot{h}(t) H_*(t) \dot{Q}_*(t) \gamma _*(t)-24 h(t) \dot{H}_*(t) \dot{Q}_*(t) \gamma _*(t)+9 h(t) \gamma _*(t){}^2 \ddot{Q}_*(t)+18 \dot{h}(t) \dot{Q}_*(t) \gamma _*(t){}^2 &\nonumber \\
     +18 h(t) \dot{Q}_*(t) \gamma _*(t) \dot{\gamma }_*(t)\Bigg)+3F^{*}_{QQQQ}(t)  \Big(9 g(t) H_*(t) \dot{Q}_*(t){}^2 \gamma _*(t)+3 \dot{g}(t) \dot{Q}_*(t){}^2 \gamma _*(t)&\nonumber \\
     -12 h(t) H_*(t) \dot{Q}_*(t){}^2 \gamma _*(t)+9 h(t) \dot{Q}_*(t){}^2 \gamma _*(t){}^2\Big)&=0 \,,\\
\label{eq: Set2_Pert_Expa_MD}
     3 (\mathrm{w} +1) h(t) \rho _*(t)+3 (\mathrm{w} +1) H_*(t) r(t)+\dot{r}(t) &= 0 \,.
\end{align}
\end{subequations}
As usual, \eqref{eq: Set2_Pert_Expa_Conn} is a dependent equation and does not introduce new physical dynamics. 

Notice here as well that, if considering the trivial background when $F^{*}(t)$ is identically zero, then the perturbed equations \eqref{eq: Set2_Pert_Expa_ED} to \eqref{eq: Set2_Pert_Expa_MD} reduce to the perturbed equations for connection set 1 given in \eqref{eq: set1expansion}, while the connection perturbed equation is identically satisfied. Hence, all the conclusions drawn about the stability of the theory also hold true for the trivial background in connection set 2.

\subsection{Matter dominated case} \label{Matter dominated era}
Let us first consider a case when matter energy dominates over all the other components of our Universe, or also known as nonrelativistic or matter/dust dominated regime, where $\mathrm{w}=0$.

\subsubsection{Background Expansion \label{Sec_Set2_BG_Matter}}
In matter dominated era, using the relevant background solution \eqref{eq: bgmatter} in the expression \eqref{eq: intigratingconnection} yields after integration
\begin{align}
 \label{bgfunctionalformatter}
    F^{*}_{QQ}(t)\dot{Q}_{*}(t)=\frac{\tilde{c}_{1}}{t^{2}}.
\end{align}
Subsequently, applying the relation \eqref{eq: derivativeofbgfunction} we get
\begin{align}
\label{bgfunctionalformatterFQ}
    F^{*}_{Q}(t)=\frac{-\tilde{c}_{1}}{t}+\tilde{c}_{2}.
\end{align}
We proceed to determine the background functions $F^{*}(t)$ and $\gamma_{*}(t)$ in terms of time by replacing the previous two expressions \eqref{bgfunctionalformatter} and \eqref{bgfunctionalformatterFQ} into \eqref{eq: Set2_Bg_Expa_ED} and \eqref{eq: Set2_Bg_Expa_PR}. By simultaneously solving both equations we obtain
\begin{align} 
\label{bgF*(t)}
    F^{*}(t) = \tilde{c}_{2} Q_*(t) +\frac{3 \tilde{c}_{1} ^2 Q_*(t)^2}{16 \tilde{c}_{2}}, 
\end{align}
where
\begin{align}
 Q_{*}(t)=-\frac{8 \tilde{c}_{2}}{3 \tilde{c}_{1} t}  \,, 
\end{align}
and $\gamma_{*}(t)$ is obtained to be
\begin{align}
\label{eq:bg_gma_conn2}
    \gamma_{*}(t)= \frac{8}{9t}-\frac{4\tilde{c}_{2}}{9\tilde{c}_{1}}\,.
\end{align}
The expression \eqref{bgF*(t)} defines the background function of the nonmetricity scalar in dust dominated era, which has the same GR type solution.

We can interpret this result as follows. Only the models characterized by 
\begin{align}
\label{eq: model Q + Q2}
    F(Q) &=\alpha Q + \beta Q^2
\end{align} have the precise general relativistic matter domination regime \eqref{eq: bgmatter} as a particular solution in a nontrivial manner. For this solution the integration constants are fixed as $\tilde{c}_{1}=-\sqrt{16 \alpha \beta / 3}$ and $\tilde{c}_{2}=\alpha$. Let us recall that the system of equations was 4-dimensional, c.f.\ \eqref{connectionset2F(Q)}, and the two other integration constants related to $H_*$ and $\rho_*$ where fixed in Eq.\ \eqref{eq: bgmatter}. The sign of $\tilde{c}_{1}<0$ follows from the physicality condition \eqref{eq: condition f_Q>0}, while the condition \eqref{eq: condition Q f_Q -f} implies $\tilde{c}_{2}>0$. Otherwise the solution would not be able to sustain a matter dominated era but encounters a singularity at $t>0$ different from the Big Bang. This constrains the model parameters to $\alpha>0$, $\beta>0$.

Alternatively, if the obtained background function of nonmetricity is expressed in time,
\begin{align}
    F^{*}(t) = \frac{4 \tilde{c}_{2}}{3 t^2}-\frac{8 \tilde{c}_{2}^2}{3 \tilde{c}_{1} t},
\end{align}
the subsequent derivative of the function w.r.t. $Q_{*}(t)$ can be efficiently computed using the following chain rule
\begin{align} 
\label{chainrule}
    \frac{d}{dt}(F^{*}(t)) &= \frac{dF^{*}(t)}{dQ_{*}(t)}\frac{dQ_{*}(t)}{dt} = F^{*}_{Q}(t) \dot{Q}_{*}(t), \qquad \frac{d}{dt}(F^{*}_{Q}(t)) =  F^{*}_{QQ}(t) \dot{Q}_{*}(t)\,. 
\end{align}
In either approach, the results remain the same. 
Now, by inserting equation \eqref{bgF*(t)} in equation \eqref{chainrule}, we can readily derive the subsequent background functional derivatives with respect to the nonmetricity. Afterward, substituting these values back into our functional expansion \eqref{function1expansion1}-\eqref{function1expansion4}, allows us to obtain the function $F(Q(t))$.
At this point, we are in a position to examine the stability of the standard dust dominated regime near the general relativity limit. 
\subsubsection{Perturbed expansion}

We substitute the background quantities from Eqs.\ \eqref{eq: bgmatter}, \eqref{bgfunctionalformatter}, \eqref{bgfunctionalformatterFQ}, \eqref{bgF*(t)} and \eqref{chainrule} into Eqs.\ \eqref{eq: Set2_Pert_Expa_ED}-\eqref{eq: Set2_Pert_Expa_MD}. We then retain terms only up to first order in the small parameters.  This process yields three unknown functions, $h(t)$, $r(t)$ and $g(t)$, along with three independent equations from Eqs.\ \eqref{eq: set2energydensityafterexpan}-   \eqref{set2mattercontinuityafterexpan}. After further simplification, we arrive at the following equations: 
\begin{subequations}
\label{eq: set2afterexpansion}
\begin{align}
\label{eq: set2energydensityafterexpan}
    \frac{9 \tilde{c}_{1} ^2 \dot{g}(t)}{\tilde{c}_{2} t^2 }-\frac{3 \tilde{c}_{1}  (\tilde{c}_{2} t-2 \tilde{c}_{1} ) \ddot{g}(t)}{2 \tilde{c}_{2} t}+\frac{4 h(t) ((\tilde{c}_{2}+2) t-3 \tilde{c}_{1} )}{t^2}+\frac{2 \dot{h}(t) (\tilde{c}_{2} t-2 \tilde{c}_{1} )}{t}+\frac{12 \tilde{c}_{1}  g(t)}{t^2}-2 \kappa ^2 r(t) = 0\,,  \\
\label{eq: set2pressureafterexpan}
    -\frac{4 h(t) \left(\left(\tilde{c}_{2}+2\right) t-\tilde{c}_{1} \right)}{t^2}-\frac{2 \dot{h}(t) \left(\left(\tilde{c}_{2}+2\right) t-2 \tilde{c}_{1} \right)}{t}-\frac{3}{2} \tilde{c}_{1}  \ddot{g}(t)-\frac{6 \tilde{c}_{1}  \dot{g}(t)}{t} = 0 \,,\\
\label{set2connectionafterexpan}
     -\frac{3 \tilde{c}_{1}  \dddot{g}(t) \left(\tilde{c}_{2} t-2 \tilde{c}_{1} \right)}{2 \tilde{c}_{2} t}-\frac{6 \tilde{c}_{1}  \left(\tilde{c}_{2} t-2 \tilde{c}_{1} \right) \ddot{g}(t)}{\tilde{c}_{2} t^2}+\frac{2 \left(\tilde{c}_{2} t-2 \tilde{c}_{1} \right) \ddot{h}(t)}{t}+\frac{4 \dot{h}(t) \left(\tilde{c}_{2} t-2 \tilde{c}_{1} \right)}{t^2}-\frac{4 h(t) \left(\tilde{c}_{2} t-2 \tilde{c}_{1} \right)}{t^3} = 0 \,,\\
\label{set2mattercontinuityafterexpan}
     \frac{4 h(t)}{\kappa ^2 t^2}+\frac{2 r(t)}{t}+\dot{r}(t) = 0 \,.
\end{align}
\end{subequations}
It is worth noting that our above set of equations does not include the small perturbed term $q(t)$. As explained in Sec.\ \ref{Perturbed Expansion}, we demonstrated how to obtain it as \eqref{eq: set2smallq(t)} by using the background $Q_{*}(t)$ and the perturbed expansion of $Q(t)$. 

There are various methods to solve this combined set of equations \eqref{eq: set2afterexpansion} to determine the behavior of small perturbations $r(t)$, $h(t)$, and $g(t)$. The complexity of the governing differential equation \eqref{eq: set2afterexpansion} in our case inhibits a straightforward analytical solution. However, alternative approaches can be employed to make progress. Dynamical system analysis, power law ansatz, and other techniques offer possibilities for tackling such equations. In our analysis, we carefully found a power law ansatz that captures all the essential features of the solution that solves the system completely. This ansatz takes the following form
\begin{equation}
 \label{ansatz}
  r(t) = \sum_{n}^{}\tilde{r}_{n}\ t^{n},   \qquad  h(t) = \sum_{n}^{}\tilde{h}_{n}\ t^{n}, \qquad g(t) = \sum_{n}^{}\tilde{g}_{n}\ t^{n} \,,
\end{equation}
where $\tilde{r}_n$, $\tilde{h}_n$, and $\tilde{g}_n$ are constants and $n$ are integers running from negative infinity to infinity. Now, by plugging the ansatz \eqref{ansatz} into Eq.\ \eqref{eq: set2afterexpansion}, we solve for all the coefficients in front of $t^n$ in such a way that they are compatible with all the equations. After a careful analysis, the nonzero contributions to the perturbations turn out to be
\begin{align}
\label{Perturbed_solution_matter_dominated}
    r(t) = \frac{\tilde{r}_2}{t^2}+\frac{4 \tilde{h}_2}{\kappa ^2 t^3}, \qquad h(t) = \frac{\tilde{h}_2}{t^2}, \qquad g(t) = \frac{\tilde{r}_2 \kappa ^2}{6 \tilde{c}_{1} }+\frac{4 \tilde{h}_2}{3 t^2}.
\end{align}
Recall the integration constants $\tilde{c}_{1}$ and $\tilde{c}_{2}$ were introduced in solving the background equations for matter dominated era. Our analysis demonstrates that small perturbations, $h(t)$ and $r(t)$, exhibit asymptotic decay over time, signifying the stability of Hubble expansion and matter energy density. However, the marginal stability of the background connection raises concerns about the long term stability of a matter dominated era around the general relativity limit.
In Sec.\ \ref{Numerical example}, we investigated further, through a numerical example, whether the marginally stable behavior in the solution could be potentially harmful. 

\subsection{Radiation dominated era} \label{Radiation dominated era}
The regime of radiation domination refers to a specific era in the early universe when radiation plays a crucial role in shaping the dynamics of the universe, and can be described through a perfect fluid with equation of state $\mathrm{w}=\frac{1}{3}$.

\subsubsection{Background evolution \label{Sec_Set2_BG_Rad}}
For a radiation dominated era, using the background $H_{*}(t)$ from \eqref{eq: bgradiation} in \eqref{eq: intigratingconnection} grants us the following functional forms
\begin{align}
    F^{*}_{QQ}(t)\dot{Q}_{*}(t) = \frac{\breve{c}_{1}} {t^{\frac{3}{2}}} , \qquad F^{*}_{Q}(t) =  \breve{c}_{2}- \frac{2 \breve{c}_{1}}{\sqrt{t}}.
\end{align}
By utilizing \eqref{eq: Set2_Pert_Expa_ED} and \eqref{eq: Set2_Pert_Expa_PR}, the background function can be expressed as 
\begin{align}\label{rad_background_fandgma}
    F^{*}(t) = \frac{3 \breve{c}_{2}}{2 t^2}-\frac{\breve{c}_{2}^2}{\breve{c}_{1} t^{3/2}}, 
\end{align}
and the background connection function becomes
\begin{align}
     \gamma_{*}(t) = \frac{1}{t}-\frac{\breve{c}_{2}}{3 \breve{c}_{1} \sqrt{t}}.
\end{align}
We can also reformulate the aforementioned function
$F_{*}(t)$ \eqref{rad_background_fandgma} in terms of the background value of the nonmetricity scalar, which looks like
\begin{align}
   \label{set2_Rad_BG_Functional}
     F^{*}(t) = Q_*(t) \left(\breve{c}_{2}-\frac{3}{2} \breve{c}_{1} \sqrt[3]{-\frac{\breve{c}_{1} Q_*(t)}{\breve{c}_{2}}}\right),
\end{align}
where 
\begin{align}
    Q_{*}(t)= -\frac{\breve{c}_{2}}{\breve{c}_{1} t^{3/2}}.
\end{align}
Just like in the matter dominated case, we have substituted the background function \eqref{rad_background_fandgma} into the chain rule \eqref{chainrule} for obtaining the  subsequent derivatives of the function with respect to the nonmetricity scalar as follows
\begin{align}
 \label{rad_bg_functional}
 F_{Q}^{*}(t) = \breve{c}_{2}-\frac{2 \breve{c}_{1}}{\sqrt{t}} , \qquad F_{QQ}^{*}(t) = \frac{2 \breve{c}_{1}^2 t}{3 \breve{c}_{2}}.
\end{align}
Hence we conclude that, unlike the connection set 1, the connection set 2 demonstrates that a single function $F_{*}(t)$ cannot simultaneously accommodate both the matter and radiation eras. In the matter era, we obtained a function with a quadratic correction in $Q_{*}(t)$ \eqref{bgF*(t)}, while the radiation era requires a function with a cubic root correction in $Q_{*}(t)$ \eqref{set2_Rad_BG_Functional}.

\subsubsection{Perturbed expansion}
We substitute the solution expansion \eqref{solexpansion}, functional expansion \eqref{functionalexpansion}, perturbed function $q(t)$ \eqref{eq: set2smallq(t)}, Hubble expansion \eqref{eq: bgradiation}, and functional derivative of nonmetricity \eqref{rad_bg_functional} into \eqref{eq: Set2_Pert_Expansion}. This substitution yields the following equations
\begin{subequations}
\label{eq: set2_rad_perturbed}
    \begin{align}
\label{eq: rad_perturbed_Ed}
    \left(-\frac{16 \breve{c}_{1}}{t^{3/2}}+\frac{3 \breve{c}_{1}^2}{\breve{c}_{2} t^2}+\frac{5 \breve{c}_{2}}{t}+\frac{6}{t}\right) h(t)+\left(\frac{6 \breve{c}_{1}^2}{\breve{c}_{2} t}-\frac{8 \breve{c}_{1}}{\sqrt{t}}+2 \breve{c}_{2}\right) \dot{h}(t)+\left(\frac{6 \breve{c}_{1}}{t^{3/2}}+\frac{9 {\breve{c}_{1}}^2}{2 \breve{c}_{2} t^2}\right) g(t) +\left(\frac{18 {\breve{c}_{1}}^2}{\breve{c}_{2} t}-\frac{3 \breve{c}_{1}}{\sqrt{t}}\right) \dot{g}(t) &\nonumber \\
    +\left(\frac{6 {\breve{c}_{1}}^2}{\breve{c}_{2}}-2 \breve{c}_{1} \sqrt{t}\right) \ddot{g}(t) & = 2 \kappa ^2 r(t)\,,\\
\label{eq: rad_perturbed_Pr}  
    \left(\frac{4 \breve{c}_{1}}{t^{3/2}}+\frac{\breve{c}_{1}^2}{\breve{c}_{2} t^2}-\frac{3 \breve{c}_{2}}{t}-\frac{6}{t}\right) h(t)+\left(\frac{2 \breve{c}_{1}^2}{\breve{c}_{2} t}+\frac{4 \breve{c}_{1}}{\sqrt{t}}-2 \breve{c}_{2}-4\right) \dot{h}(t) + \frac{3 \breve{c}_{1}^2 g(t)}{2 \breve{c}_{2} t^2}  +\left(\frac{6 \breve{c}_{1}^2}{\breve{c}_{2} t}-\frac{7 \breve{c}_{1}}{\sqrt{t}}\right) \dot{g}(t)   &\nonumber\\
    +\left(\frac{2 \breve{c}_{1}^2}{\breve{c}_{2}}-2 \breve{c}_{1} \sqrt{t}\right) \ddot{g}(t) & = \frac{2}{3} \kappa ^2 r(t)\,,  \\
\label{eq: rad_perturbed_Conn} 
    \left(\frac{6 \breve{c}_{1}}{t^{5/2}}-\frac{2 \breve{c}_{2}}{t^2}\right) h(t)+\left(-\frac{18 \breve{c}_{1}}{t^{3/2}}+\frac{9 \breve{c}_{1}^2}{\breve{c}_{2} t^2}+\frac{5 \breve{c}_{2}}{t}\right) \dot{h}(t)+\left(\frac{6 \breve{c}_{1}^2}{\breve{c}_{2} t}-\frac{8 \breve{c}_{1}}{\sqrt{t}}+2 \breve{c}_{2}\right) \ddot{h}(t)+\left(\frac{45 \breve{c}_{1}^2}{2 \breve{c}_{2} t^2}-\frac{15 \breve{c}_{1}}{2 t^{3/2}}\right) \dot{g}(t)  &\nonumber \\
    +\left(\frac{30 \breve{c}_{1}^2}{\breve{c}_{2} t}-\frac{10 \breve{c}_{1}}{\sqrt{t}}\right) \ddot{g}(t)+\left(\frac{6 \breve{c}_{1}^2}{\breve{c}_{2}}-2 \breve{c}_{1} \sqrt{t}\right) \dddot{g}(t)  
    &=0\,,\\
\label{eq: rad_perturbed_Md} 
    \frac{3 h(t)}{\kappa ^2 t^2}+\frac{2 r(t)}{t}+\dot{r}(t) & =0\,.
    \end{align}
\end{subequations}
 We aim to understand the evolutionary behavior of the small perturbed function $r(t)$, $h(t)$ and $g(t)$ by solving the combined differential Eqs.\ \eqref{eq: set2_rad_perturbed}. Building upon the success in the matter dominated era case, we adopt a similar strategy for the radiation dominated era. To proceed, we choose an ansatz that encompasses all the entire solution space for this epoch,
\begin{equation}
\label{set2_rad_ansatz}
    r(t) = \sum_{n}^{} \breve{r}_{n}\ t^{\frac{n}{2}}, \qquad 
    h(t) = \sum_{n}^{} \breve{h}_{n}\ t^{\frac{n}{2}}, \qquad 
    g(t) = \sum_{n}^{} \breve{g}_{n}\ t^{\frac{n}{2}} \,,
\end{equation}
where $\breve{r}_n$, $\breve{h}_n$, and $\breve{g}_n$ are constants. As noted previously, $\breve{c}_{1}$ and $\breve{c}_{2}$ are integration constants that emerged during the solution of the background equations for the radiation dominated era. Substituting our ansatz \eqref{set2_rad_ansatz} into Eq.\ \eqref{eq: set2_rad_perturbed}, we obtain the following for the small perturbations
\begin{align}
    r(t)= \frac{\breve{r}_4}{t^2} + \frac{3 \breve{h}_4}{\kappa ^2 t^3}\,, \qquad h(t)= \frac{\breve{g}_4}{2 t^2}\,, \qquad g(t)=  \left(\frac{2}{t^2}-\frac{\breve{c}_{2}}{3 \breve{c}_{1} t^{3/2}}\right)\breve{h}_4+\frac{\breve{g}_1}{\sqrt{t}}\, ,
\end{align}
with
\begin{align}
    \breve{h}_{4}=\frac{\breve{g}_{4}}{2} \,, \qquad \breve{r}_{4}= \frac{3\breve{c}_{1}\breve{g}_{1}}{\kappa^{2}}.
\end{align}
Different from the result for matter dominated era, the radiation dominated universe exhibits asymptotic decay over time for all small perturbations $h(t)$, $r(t)$, and $g(t)$. This behavior suggests the possibility of a stable radiation dominated universe in the vicinity of the general relativity limit.

\subsection{Dark energy dominated era} \label{Dark energy dominated era}
Our current Universe is characterized by dark energy density, while the contributions from matter and radiation densities are negligible. In this regime, we choose $\mathrm{w}=-1$ and $\rho_{*}(t)=$  constant. We investigate a nonmetricity background function suitable for the dark energy dominated era, demonstrating that it possesses a solution equivalent to GR. Furthermore, we analyze the behavior of the perturbed function to determine whether it possesses a stable dark energy dominated era or not.
 
\subsubsection{Background evolution} \label{Sec_Set2_BG_DE}

It turns out that, when the Hubble function is constant \eqref{eq: bgdarkenergy}, the solution of Eq.\ \eqref{eq: intigratingconnection} becomes incompatible with the Friedmann equations \eqref{eq: Set2_Bg_Expa_ED}-\eqref{eq: Set2_Bg_Expa_PR}, which prevents us from solving the cosmological equation as we did in dust and radiation cases. Our objective is to identify a function $F_{*}(t)$ that solves the cosmological equations \eqref{eq: Set2_Bg_Expansion}. After substituting the dark energy Hubble expansion \eqref{eq: bgdarkenergy}, the background Friedmann equations can be expressed as
\begin{subequations}
\label{eq: bg_de_subhub}
    \begin{align}
        F^*(t)+F_Q^*(t) \left(4 \Lambda_{DE} \kappa ^2-3 \sqrt{3\Lambda_{DE}}  \kappa  \gamma _*(t)-3 \dot{\gamma }_*(t)\right)+9 \gamma _*(t) F_{\text{QQ}}^*(t) \left(\sqrt{3\Lambda_{DE}}  \kappa  \dot{\gamma }_*(t)+\ddot{\gamma }_*(t)\right)&=0\,,\\
        F^*(t)+F_Q^*(t) \left(4 \Lambda_{DE} \kappa ^2-3 \sqrt{3\Lambda_{DE}} \kappa  \gamma _*(t)-3 \dot{\gamma }_*(t)\right)+F_{\text{QQ}}^*(t) \left(4 \sqrt{3\Lambda_{DE}}  \kappa -9 \gamma _*(t)\right) \left(\sqrt{3\Lambda_{DE}}  \kappa  \dot{\gamma }_*(t)+\ddot{\gamma }_*(t)\right)&=0\,.
    \end{align}
\end{subequations}
Together, by solving equations \eqref{eq: bg_de_subhub}, we obtained two different solutions for $\gamma_{*}(t)$, which are 
\begin{align}
\label{Set2_DE_BG_Sol}
    \gamma _*(t)=\hat{c}_{2}-\frac{\hat{c}_{1} e^{-3H_{*} t}}{3 H_*}\,, 
\end{align}
and    
\begin{align}
\label{Set2_DE_BG_Sec_Sol}
    \gamma _*(t)=\frac{2 H_*}{3}\,.
\end{align}
The function $F^*(t)$ of the nonmetricity scalar corresponding to the first solution \eqref{Set2_DE_BG_Sol} is
\begin{align}
\label{Set2_DE_BG_Fun_Sol}
    F^*(t)=\hat{c}_{3} \left(Q_*(t)-2 \Lambda_{DE} \kappa ^2\right)\,,
\end{align}
where
\begin{align}
    Q_*(t) = 3 \sqrt{3\hat{c}_{1}} \hat{c}_{2} \kappa -2 \hat{c}_{1} \kappa ^2.
\end{align}
Meanwhile, for the second solution \eqref{Set2_DE_BG_Sec_Sol} we obtain that the function is
\begin{align}
\label{Set2_DE_BG_Fun_Sec_Sol}
    F^*(t) = \hat{c}_{4} e^{-\frac{Q_*(t)}{2 \Lambda_{DE} \kappa ^2}},
\end{align}
where
\begin{align}
\label{eq:2nd_bra_nonmet}
    Q_*(t)=0.
\end{align}

Due to the linearity of the function \eqref{Set2_DE_BG_Fun_Sol} in the  nonmetricity scalar, the higher order derivative term vanishes, which simplifies the perturbed equation a lot, as we will show in the next subsection.

\subsubsection{Perturbed expansion}
Substituting the background evolution \eqref{Set2_DE_BG_Sol}-\eqref{Set2_DE_BG_Fun_Sol} to the perturbed expansion Eqs.\ \eqref{eq: Set2_Pert_Expa_ED}-\eqref{eq: Set2_Pert_Expa_MD} yields
\begin{subequations}
\label{eq: Set2_DE_Pert}
     \begin{align}
\label{eq: Set2_DE_Pert_ED}
         2 \kappa  \left(2 \sqrt{3\Lambda_{DE}}  \left(\hat{c}_{3}+1\right) h(t)-\hat{c}_{5} \kappa \right) &=0 \,,\\
\label{eq: Set2_DE_Pert_PR}
         2 \hat{c}_{5} \kappa ^2-4 \sqrt{3\Lambda_{DE}}  \left(\hat{c}_{3}+1\right) \kappa  h(t)-4 \left(\hat{c}_{3}+1\right) \dot{h}(t) &=0 \,,
     \end{align}
 \end{subequations}
while the connection equation \eqref{eq: Set2_Pert_Expa_Conn} remains identically satisfied.
Solving equations \eqref{eq: Set2_DE_Pert_ED} and \eqref{eq: Set2_DE_Pert_PR} allows us to readily obtain the perturbed solution as
 \begin{align}
    r(t) \sim \hat{c}_{5}\,, \qquad  h(t) \sim \hat{c}_{6}\,.
 \end{align}
Considering the second solution \eqref{Set2_DE_BG_Sec_Sol} and its associated nonmetricity function \eqref{Set2_DE_BG_Fun_Sec_Sol}, the resulting perturbed equations become
\begin{subequations}
\label{eq: Set2_DE_Pert_sec}
     \begin{align}
\label{eq: Set2_DE_Pert_ED_sec}
        \frac{\sqrt{3} \hat{c}_{4} \ddot{g}(t)}{2 c_1^{3/2} \kappa ^3}+\frac{3 \hat{c}_{4} \dot{g}(t)}{\Lambda_{DE} \kappa ^2}+\frac{3 \sqrt{3} \hat{c}_{4} g(t)}{2 \sqrt{\Lambda_{DE}} \kappa }-\frac{\hat{c}_{4} \dot{h}(t)}{\Lambda_{DE} \kappa ^2}+\left(4 \sqrt{3} \sqrt{\Lambda_{DE}} \kappa -\frac{3 \sqrt{3} \hat{c}_{4}}{\sqrt{\Lambda_{DE}} \kappa }\right) h(t)-2 \hat{c}_{5} \kappa ^2 &= 0\,,\\
\label{eq: Set2_DE_Pert_PR_sec} 
        -\frac{\sqrt{3} \hat{c}_{4} \ddot{g}(t)}{2 \Lambda_{DE}^{3/2} \kappa ^3}-\frac{3 \hat{c}_{4} \dot{g}(t)}{\Lambda_{DE} \kappa ^2}-\frac{3 \sqrt{3} \hat{c}_{4} g(t)}{2 \sqrt{\Lambda_{DE}} \kappa }+\frac{3 \hat{c}_{4} \dot{h}(t)}{\Lambda_{DE} \kappa ^2}+\left(\frac{3 \sqrt{3} \hat{c}_{4}}{\sqrt{\Lambda_{DE}} \kappa }-4 \sqrt{3} \sqrt{\Lambda_{DE}} \kappa \right) h(t)+2 \hat{c}_{5} \kappa ^2-4 \dot{h}(t)  &= 0\,,\\
\label{eq: Set2_DE_Pert_Conn_sec}
        \frac{\sqrt{3} \hat{c}_{4} \dddot{g}(t)}{2 \Lambda_{DE}^{3/2} \kappa ^3}+\frac{3 \hat{c}_{4} \ddot{g}(t)}{\Lambda_{DE} \kappa ^2}+\frac{3 \sqrt{3} \hat{c}_{4} \dot{g}(t)}{2 \sqrt{\Lambda_{DE}} \kappa }-\frac{\hat{c}_{4} \ddot{h}(t)}{\Lambda_{DE} \kappa ^2}-\frac{\sqrt{3} \hat{c}_{4} \dot{h}(t)}{\sqrt{\Lambda_{DE}} \kappa } &= 0\,,
        \end{align}
\end{subequations}
where $\hat{c}_{1}, \ldots, \hat{c}_{8}$ denote integration constants. The solutions for the perturbed functions obtained after solving \eqref{eq: Set2_DE_Pert_sec} are expressed as
\begin{align}
\label{eq:2nd_bra_de_pert}
    h(t)\sim \hat{c}_{7},\qquad g(t)\sim \hat{c}_{8}.
\end{align}

Analyzing both solution branches reveals that neither the Hubble parameter nor the matter energy density exhibit decay or growth with time. However, the behavior of the perturbed connection function $g(t)$ differs in both branches \eqref{Set2_DE_BG_Sol} and \eqref{Set2_DE_BG_Sec_Sol} of $\gamma_{*}(t)$. In the first branch \eqref{Set2_DE_BG_Sol} we observe that the perturbed connection function $g(t)$ within Eqs.\ \eqref{eq: Set2_Pert_Expa_ED}-\eqref{eq: Set2_Pert_Expa_Conn} is multiplied by the higher  derivatives of the function \eqref{Set2_DE_BG_Fun_Sol}, thus these terms vanish due to the linearity of the function  \eqref{Set2_DE_BG_Fun_Sol}. Hence, the connection perturbation $g(t)$ remains undetermined.
Conversely, in the second branch \eqref{Set2_DE_BG_Sec_Sol} although the nonmetricity scalar \eqref{eq:2nd_bra_nonmet} becomes zero, the function \eqref{Set2_DE_BG_Fun_Sec_Sol} and its derivative evolve as constant, and that allows us to determine the connection perturbation \eqref{eq:2nd_bra_de_pert}.
As a result, stability remains unclear in the first branch of $\gamma_{*}(t)$, while the second branch suggests marginal stability when approaching the limit of general relativity.

\subsection{Geometric dark energy} \label{Geometric dark energy dominated era}
For the geometric dark energy regime, we assume a background matter energy density to be zero, i.e., $\rho_{*}(t)=0$ while retaining dust perturbations $r(t)$ with an equation of state $\mathrm{w}=0$.  As we learned, the background function obtained from the previous section will no longer be valid here. In this section, we will analyze the respective background function of the geometric dark energy era for which the field equations reduce to GR. We will subsequently analyze the behavior of perturbed functions to investigate the possibility of a stable era near the GR limit. 
\subsubsection{Background evolution} \label{Sec_Set2_BG_GDE}
By comparing the background Friedmann equation expressed in \eqref{eq: Set2_Bg_Expa_ED}--\eqref{eq: Set2_Bg_Expa_PR} with the standard general relativity Friedmann equation \eqref{eq: bg_gddark}, we obtained the following additional contribution from GDE:
\begin{subequations}
\label{eq: Set2_GDE_BG}
    \begin{align}
    \label{eq: Set2_GDE_BG_ED}
         -\big(\left(6 H_*(t){}^2 -Q_*(t)\right) F^{*}_Q(t)+3 \dot{Q}_{*}(t) \gamma _*(t) F^{*}_{QQ}(t)+F^*(t)\big)& = 2\kappa^{2}\Lambda_{GDE}\,,\\
    \label{eq: Set2_GDE_BG_PR}
         F^{*}_Q(t)\left(6 H_*(t){}^2+4 \dot{H}_*(t)-Q_*(t)\right) +F^{*}_{QQ}(t) \left(4 H_*(t) \dot{Q}_*(t)-3 \dot{Q}_*(t) \gamma _*(t)\right)+F^*(t) & =  2\kappa^{2}\Lambda_{GDE}\,.
    \end{align}
The nGR part of Eqs.\ \eqref{eq: Set2_GDE_BG_ED}-\eqref{eq: Set2_GDE_BG_PR} plays the role of the cosmological constant or dark energy. The background connection equation is represented by Eq.\ \eqref{eq: Set2_Bg_Expa_Conn}, while the background matter continuity equation is identically satisfied.
\end{subequations}
For the geometric dark energy dominated era, we know that the background Hubble expansion \eqref{eq: bggeomdarkenergy} is constant. We employ the same strategy to determine the function of nonmetricity, which solves the cosmological equations of the geometric dark energy dominated era. In addition, to solve the background equation consistently, we assume the time derivative of the nonmetricity scalar should vanish. Now substituting the Hubble parameter \eqref{eq: bggeomdarkenergy} along with assumption into the background Friedmann equations \eqref{eq: Set2_GDE_BG} and together solving the equations, we can easily solve for $\gamma_{*}$ as 
\begin{align}
\label{set2_GDE_gamma}
    \gamma _*(t) = \frac{e^{-3 H_* t}\bar{c}_{1}}{3 H_*} + \bar{c}_{2} \
\end{align}
where $\bar{c}_{1}$ to $\bar{c}_{4}$ denote integration constants. With this, the function of nonmetricity takes the form 
\begin{align}
\label{set2_GDE_BG_Functional}
    F^*(t)=2 \left(3 \kappa ^2 \Lambda_{GDE} -2 Q_*(t)\right)\,,
\end{align}
which is linear in the nonmetricity scalar, and where the latter is given by
\begin{align}
    Q_*(t) = -2 \kappa ^2 \Lambda_{GDE} +3 \sqrt{3\Lambda_{GDE}} \bar{c}_{2} \kappa.
\end{align}

\subsubsection{Perturbed expansion}
Having established all the background evolution, we move forward to obtain the perturbation functions. Now, substituting the background evolution \eqref{eq: bggeomdarkenergy}, \eqref{set2_GDE_gamma} and \eqref{set2_GDE_BG_Functional} to the perturbed expansion Eqs.\ \eqref{eq: Set2_Pert_Expa_ED}-\eqref{eq: Set2_Pert_Expa_MD}, it is obtained
\begin{subequations}
\label{eq: Set2_GDE_Pert}
    \begin{align}
\label{eq: Set2_GDE_Pert_ED}
        -2\left(6 \sqrt{3\kappa^{2}\Lambda_{GDE}}   h(t) + \kappa ^2 r(t)\right)=0\,,\\
\label{eq: Set2_GDE_Pert_PR}
        -12 \left( \dot{h}(t)+\sqrt{3\kappa^{2}\Lambda_{GDE}}  h(t)\right) =0\,,\\     
\label{eq: Set2_GDE_Pert_MD}
        \dot{r}(t)+\sqrt{3\kappa^{2}\Lambda_{GDE}}  r(t) =0\,,
    \end{align}
\end{subequations}
while the connection equation is identically satisfied due to the linearity of $F$ in $Q$ \eqref{set2_GDE_BG_Functional} and also it significantly simplifies the remaining equations of \eqref{eq: Set2_GDE_Pert}. Notice that all perturbed equations are free from the connection perturbed function $g(t)$. Again, the same reason is that the linearity of the function \eqref{set2_GDE_BG_Functional} ensures the absence of the connection perturbation $g(t)$ from the original perturbed equations \eqref{eq: Set2_Pert_Expansion} due to vanishing higher order derivative terms of it. A straightforward solution of equations \eqref{eq: Set2_GDE_Pert} provides the perturbed Hubble function and energy density as
\begin{align}
    r(t) \sim \bar{c}_{3} e^{-3H_{*}t}\,,\qquad h(t) \sim \bar{c}_{4} e^{-3H_{*}t}\,.
\end{align}
As a result, although $h(t)$ and $r(t)$ exhibit an asymptotic decreasing behavior with time, without knowing the evolution of connection perturbed function $g(t)$, the stability of a geometric dark energy dominated era remains unclear.

\section{Numerical example} 
\label{Numerical example}
In the previous section on the connection set 2 we learned several things. First, the general relativity radiation dominated, matter dominated, and dark energy dominated era  exact solutions exist in theories which either trivially allow $F(Q_*)=0$ or for different model functions $F(Q)$ depending on the era. Second, in the latter case while the Hubble perturbations and density perturbations near the matter dominated solution converge, the perturbations of the connection function $\gamma$ remain only marginally stable. On the other hand we know from the general analysis of Sec.\ \ref{Connection set 2} that the system of equations is prone to experience sudden singularities for any model $f(Q)$, just depending on the evolution of the connection function $\gamma$. In order to get a better feel whether the marginal stability of $\gamma$ could be a serious problem or not, it will be good to get a more direct peek into the situation by taking a numerical example.

As discussed in the earlier Sec.\ \ref{Sec_Set2_BG_Matter}, the correct cosmological behavior 
in the matter dominated era with connection set 2 was consistent with the quadratic model function. Thus, let us
study the system \eqref{eq: set2 syn sys} numerically, 
by taking the function \eqref{eq: model Q + Q2} with the parameter values $\alpha=1$, $\beta=1$, and set the units with $\kappa=1$. The evolution of the system is then described by Eqs.\ \eqref{eq: set2 syn sys} with these values substituted in. We immediately notice, that the system inhabits a singularity of diverging $\dot{H}$ and $\dot{\Pi}$ when 
\begin{align}
    1 + F_Q &= - 12 H^2 + 18 H \gamma + 6 \dot{\gamma} +2
\end{align}
vanishes (cf.\ the discussion in Sec.\ \ref{sec: general conditions of stability}) or $\gamma$ vanishes. The latter would violate the premise of deriving the connection \eqref{connectionset2}, and therefore it is not a surprise that the system harbors a singularity there.

\begin{figure}
    \begin{center}
    \subfigure[]{
    \includegraphics[width=0.45\textwidth]{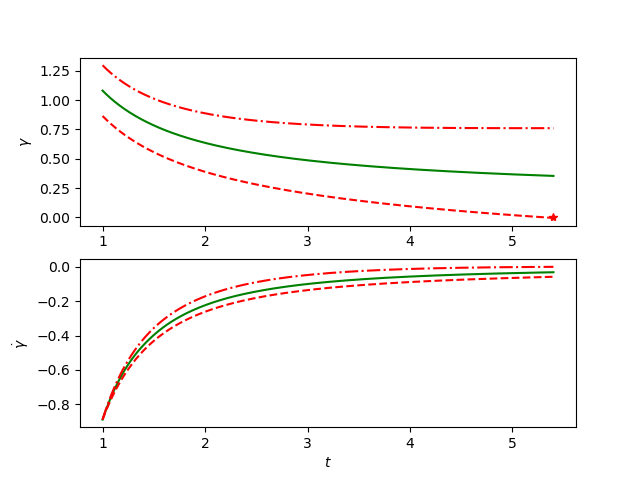}
    \label{fig: Q Q2 set2 gamma gammadot}}
    \subfigure[]{
    \includegraphics[width=0.45\textwidth]{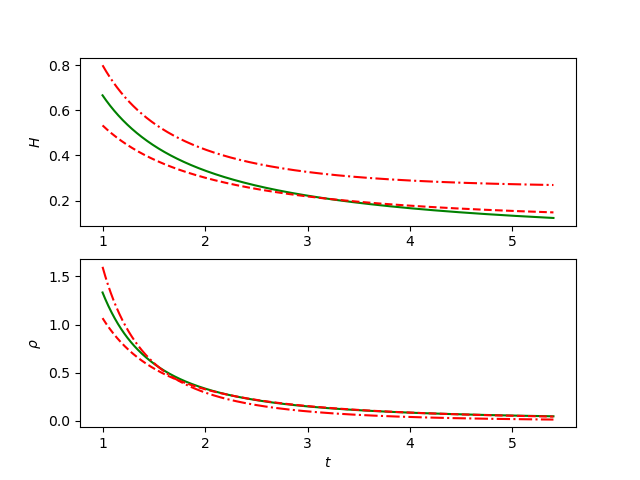}
    \label{fig: Q Q2 set2 H rho}}
    \end{center}
    \caption{Numerical solutions of the system \eqref{eq: set2 syn sys} for the model function \eqref{eq: model Q + Q2} with $\alpha=1$, $\beta=1$, $\kappa=1$. The green solid curves depict the matter dominated solution with initial conditions matching general relativity ($t=1$, $H=0.666$, $\rho=1.333$, $\gamma=1.081$, $\dot{\gamma}=-0.888$), while the red curves depict 20\% perturbed configurations (dashed-dot case $H=0.8$, $\rho=1.6$, $\gamma=1.297$; dashed case $H=0.533$, $\rho=1.066$, $\gamma=0.865$ but $\dot{\gamma}$ the same). The latter solution hits a singularity by reaching $\gamma=0$ at about $t=5.4$, indicated by the star symbol.}
    \label{fig: numerical}
\end{figure}

A simple set of initial conditions for the solution that reproduces the matter dominated regime in general relativity can be found by taking $t=1$ in \eqref{eq: bgmatter} that yields $H_*$ and $\rho_*$, and fixing $\tilde{c}_1$, $\Tilde{c}_2$ in terms of $\alpha$, $\beta$ as explained in Sec.\ \ref{Sec_Set2_BG_Matter} to obtain $\gamma_*$ and $\dot{\gamma}_*=\Pi_*$ from \eqref{bgF*(t)}. The numerical evolution of the system \eqref{eq: set2 syn sys} from these initial conditions is depicted on Fig.\ \ref{fig: numerical} by the solid green curve. As expected from the previous GR limit matter dominated solution \eqref{eq: bgmatter}, $H_*$ and $\rho_*$ decrease as first and second powers of time, while the background $\gamma_*$ obtained in Sec.\Ref{Sec_Set2_BG_Matter} also converges to a constant value.

To understand the significance of the marginal stability of the connection function, we can numerically run the equations for some other solutions in the neighborhood, starting with slightly different initial values for $\gamma$, $H$, and $\rho$. The result is depicted on Fig.\ \ref{fig: numerical} as dashed red curves. While the values of $H$ and $\rho$ of the perturbed solutions start to come closer to the respective values of the background solution as predicted by \eqref{Perturbed_solution_matter_dominated}, the value of $\gamma$ does neither approach nor strongly depart form the background value in accord with its marginal stability. Since the deviation of $\gamma$ from the background solution does not get smaller in time, the system does not force $H$ and $\rho$ to completely converge to their background values either. The solutions would be truly stable if all variables had stable behavior.

The most significant feature here is that one of those perturbed solutions hits a singularity after some time. The singularity is reached at $\gamma=0$, which implies divergent $\dot{H}$ and subsequently all other quantities will destabilize. We have not continued the solution into the singularity on the plots, because that would need specialized numerical techniques to provide a reliable result. However, this quick example demonstrates that for a large class of initial conditions that are not too far from the good general relativity type behavior, the solution evolves into a sudden singularity without an obvious hindrance. Hence on top of the difficulty in encompassing all major cosmological eras into a single $f(Q)$ model, even in the model where the set 2 connection matter dominated solution exists, it is not really stable as deviations from that solution can end up in a singularity.

\section{Conclusions} \label{Conclusions}
In this work, we studied the stability of cosmological solutions corresponding to different connections in symmetric teleparallel $f(Q)$ gravity. 
First by inspecting the field equations, we outlined several conditions which must be obeyed by an $f(Q)$ model that could be considered relevant from a physical point of view. 
Let us emphasize that in general terms it can not be said that a certain model $f(Q)$ is stable or a certain connection is stable, but these two aspects must be considered together. It could happen that some particular connection is well behaved in conjunction with certain class of models, but not in conjunction with some others, while conversely the latter could still yield agreeable physics when utilizing them with a different connection.

These considerations become especially relevant in cosmology, where it is known that the spatially homogeneous and isotropic connections that satisfy the symmetric teleparallel conditions paired with a flat Friedmann-Lema\^itre-Robertson-Walker metric can be classified into four sets. Three of these branches correspond to a cosmology with zero spatial curvature $(k=0)$, while the last branch corresponds to curvature $k\neq 0$. We confined our work only to the first and second branches of the connection. The first is the trivial case which has been extensively studied in diverse contexts, while the second introduces an additional free function $\gamma$ in the cosmological equations, and has been recently reported to contain a ghost degree of freedom in the ultraviolet regime. 

We focused our study on the behavior near the general relativity $\Lambda$CDM limit, since the solutions that can observationally describe our Universe are most likely to be found in this neighborhood. In principle there are three options how to reach the general relativity limit. The first is realized trivially when the model function $f(Q)$ has a linear regime that approximates $Q$ (plus a cosmological constant). The second occurs also trivially when the connection evolves in a manner whereby the nonmetricity scalar $Q$ is constant or zero. However, this option is quite restricted if we assume the connection obeys FLRW symmetry, as these two conditions together select specific Hubble functions only. The third option to obtain the general relativity limit happen by a nontrivial cancellation among the terms in the equations of motion. We have found that in the connection set 1 case, the second option only selects Minkowski spacetime, while the third option picks out a linear plus square root regime \eqref{eq: set1 F}. It turns out that the perturbations around the linear $f(Q)$ regime and around the nontrivial regime have the same stability properties, namely they are stable in the matter, radiation, and geometric dark energy dominated eras, and a marginally stable in the dark energy dominated era.
In the case of connection set 2, all three options to realize the general relativity limit are available, but in contrast to the set 1 case the constant $Q$ implies de Sitter, while the nontrivial cancellation picks out different functional regimes of different powers of $Q$ for different eras. The properties of the perturbations in the trivial regimes coincide with the ones of set 1.
In the dust dominated era, the nontrivial GR regime corresponds to a quadratic correction term in $f(Q)$, and the connection 
function is marginally stable around the GR solution. In the radiation era, the GR regime is achieved with a cubic correction in $f(Q)$, and all perturbed functions are stable. We found two possible solutions for the connection function $\gamma_{*}(t)$ in the dark energy era, where in the first case its stability can not be determined while in the second it is marginally stable. Finally, in the geometric dark energy era the stability remains undetermined as well.

We may consider a model of $f(Q)$ viable in the cosmological sense if it can approximate the $\Lambda$CDM scenario and the perturbations of the initial conditions around this base scenario are stable, i.e.\ the dynamics does not get wildly diverted away from it. In this respect, therefore both set 1 and 2 connections offer possibilities for a realistic description of the Universe, provided the model function $f(Q)$ is suitable. Only the nontrivial realization of the GR regime in set 2 has probably less applications because different eras imply different behavior of $f(Q)$ and the stability is only guaranteed in the radiation domination epoch.

The study of stability is also relevant for the understanding of how susceptible are the cosmological solutions to a sudden singularity where $\dot{H}$ diverges at finite matter density. Like in $f(R)$ gravity, there definitely are models among the $f(Q)$ family that allow such behavior. However, what really  concerned us here was whether this phenomenon can occur near the general relativity regime as well, i.e.\ in physically relevant conditions. Inspection of the field equations in the form of a dynamical system showed the conditions for this possibility. Although the inclusion of an extra function $\gamma$ in connection set 2 allows more freedom in fitting the data, it also made the system more vulnerable for a singularity to occur, as the value of $\gamma$ is not blocked from switching the sign, but which would suddenly wreck the system into a singularity. We demonstrated this explicitly happening by numerically evolving the equations from the initial conditions taken near the solution that corresponds to a dust matter universe in general relativity. 

The occurrence of sudden singularities from otherwise perfectly normal looking initial conditions in a reasonable $f(Q)$ model is a warning that cosmologies  with the alternative FLRW connection set 2 (and by analogy also set 3) must be treated with great caution. Clearly, the phenomenon of sudden singularities needs further investigation, and it could turn out that the alternative connections are unable to provide reliable backgrounds for cosmology. In this respect finding a ghost among the metric perturbations is perhaps not of great consequence if already the background is pathological.

\section*{Acknowledgments}

This work was supported by the European Regional Development Fund through the Center of Excellence TK133 ``The Dark Side of the Universe,'' and the Estonian Research Council grants PRG356 ``Gauge Gravity'' and TK202 ``Foundations of the Universe.'' M.J.G. and L.P. have been supported by the Estonian Research Council grant PSG910 ``Theoretical frameworks for numerical modified gravity''. We would like to thank all the members of the Tartu Journal Club for their insightful discussion and an anonymous referee for their valuable comments.

\end{document}